\RequirePackage{fix-cm} 
\documentclass[fleqn,usenatbib]{mnras}

\usepackage{newtxtext,newtxmath}

\usepackage[T1]{fontenc}

\DeclareRobustCommand{\VAN}[3]{#2}
\let\VANthebibliography\thebibliography
\def\thebibliography{\DeclareRobustCommand{\VAN}[3]{##3}\VANthebibliography}

\usepackage{subcaption}
\usepackage{physics}
\usepackage[per-mode=reciprocal]{siunitx} 
\usepackage[dvipsnames]{xcolor}  

\AtBeginDocument{\RenewCommandCopy\qty\SI}
\ExplSyntaxOn
\msg_redirect_name:nnn { siunitx } { physics-pkg } { none }
\ExplSyntaxOff

\usepackage{tikz}
\usetikzlibrary{shapes.geometric, arrows, fit}

\tikzstyle{startstop} = [rectangle, rounded corners, text centered, draw=black, fill=red!30]
\tikzstyle{io} = [trapezium, trapezium left angle=70, trapezium right angle=110, text centered, draw=black, fill=blue!30]
\tikzstyle{process} = [rectangle, text centered, draw=black, fill=orange!30]
\tikzstyle{decision} = [diamond, text centered, draw=black, fill=green!30]
\tikzstyle{arrow} = [thick,->,>=stealth]

\tikzstyle{condtext} = [text centered, minimum width=2em, minimum height=2em]

\usepackage[strings]{underscore} 

\usepackage{svg}
\usepackage{hyperref}

\usepackage{acro}
\DeclareAcronym{HMXB}{
	short=HMXB,
	long=high-mass X-ray binary,
}
\DeclareAcronym{LMXB}{
	short=LMXB,
	long=low-mass X-ray binary,
}

\DeclareAcronym{PZF}{
	short=PZF,
	long=proximity zone fossils,
}

\DeclareAcronym{eRASS1}{
	short=eRASS,
	long=the first eROSITA All Sky Survey,
}

\DeclareAcronym{SNR}{
	short=SNR,
	long=Supernova Remnant,
}

\DeclareAcronym{CMZ}{
	short=CMZ,
	long=Central Molecular Zone,
}

\DeclareAcronym{CR}{
	short=CR,
	long=cosmic ray,
}

\DeclareAcronym{tSZ}{
	short=tSZ,
	long=thermal Sunyaev-Zel'dovich,
}

\DeclareAcronym{SZE}{
	short=SZE,
	long=Sunyaev-Zel'dovich Effect,
}

\DeclareAcronym{MCMC}{
	short=MCMC,
	long=Markov Chain Monte Carlo,
}

\DeclareAcronym{ODE}{
	short=ODE,
	long=ordinary differential equation,
}

\DeclareAcronym{UVB}{
	short=UVB,
	long=ultraviolet background,
}

\DeclareAcronym{JIT}{
	short=JIT,
	long=just-in-time,
}

\DeclareAcronym{NFW}{
	short=NFW,
	long=Navarro-Frenk-White,
}

\DeclareAcronym{ISM}{
	short=ISM,
	long=interstellar medium,
}

\DeclareAcronym{IGM}{
	short=IGM,
	long=intergalactic medium,
}

\DeclareAcronym{ICM}{
	short=ICM,
	long=intracluster medium,
}

\DeclareAcronym{CGM}{
	short=CGM,
	long=circumgalactic medium,
}

\DeclareAcronym{CIE}{
	short=CIE,
	long=collisional ionisation equilibrium,
}

\DeclareAcronym{PCHIP}{
	short=PCHIP,
	long=Piecewise Cubic Hermite Interpolating Polynomial,
}

\DeclareAcronym{NSF}{
	short=NSF,
	long=nuclear star formation,
}

\DeclareAcronym{SDSS}{
	short=SDSS,
	long=Sloan Digital Sky Survey,
}

\DeclareAcronym{CFHT}{
	short=CFHT,
	long=Canada-French-Hawaii Telescope,
}

\DeclareAcronym{MOONS}{
	short=MOONS,
	long=Multi-Object Optical and Near-infrared Spectrograph,
}

\DeclareAcronym{4MOST}{
	short=4MOST,
	long=4-metre Multi-Object Spectroscopic Telescope,
}

\DeclareAcronym{RRL}{
	short=RRL,
	long=RR Lyrae star,
}

\DeclareAcronym{GC}{
	short=GC,
	long=galactic centre,
}

\DeclareAcronym{HB}{
	short=HB,
	long=horizontal branch,
}

\DeclareAcronym{CMD}{
	short=CMD,
	long=colour-magnitude diagram,
}

\DeclareAcronym{RA}{
	short=RA,
	alt=$\alpha$,
	long=right ascension,
}

\DeclareAcronym{DEC}{
	short=Dec,
	alt=$\delta$,
	long=declination,
}

\DeclareAcronym{GSR}{
	short=GSR,
	long=galactic standard of rest,
}

\DeclareAcronym{LSR}{
	short=LSR,
	long=local standard of rest,
}

\DeclareAcronym{CMB}{
	short=CMB,
	long=Cosmic Microwave Background,
}

\DeclareAcronym{SFR}{
	short=SFR,
	long=star formation rate,
}

\DeclareAcronym{AGB}{
	short=AGB,
	long=asymptotic giant branch,
}

\DeclareAcronym{SNII}{
	short=SNII,
	long=type II supernova,
}

\DeclareAcronym{SNIA}{
	short=SNIa,
	long=type Ia supernova,
}

\DeclareAcronym{IMF}{
	short=IMF,
	long=initial mass function,
}

\DeclareAcronym{SSP}{
	short=SSP,
	long=simple stellar population,
}

\DeclareAcronym{DMO}{
	short=DMO,
	long=dark-matter only,
}

\DeclareAcronym{BP}{
	short=B/P,
	long=Boxy/Peanut,
}

\DeclareAcronym{PIGS}{
	short=PIGS,
	long=Pristine Inner Galactic Survey,
}

\DeclareAcronym{MW}{
	short=MW,
	long=Milky Way,
}

\DeclareAcronym{CLB}{
	short=ClB,
	long=Classical Bulge,
}

\DeclareAcronym{PSB}{
	short=PsB,
	long=Pseudobulge,
}

\DeclareAcronym{DR2}{
	short=DR2,
	long=Gaia data release 2,
}

\DeclareAcronym{MHD}{
	short=MHD,
	long=magneto-hydrodynamical,
}

\DeclareAcronym{AGN}{
	short=AGN,
	long=active galactic nucleus,
}

\newcommand{\sgrAs}{Sgr A$^*$}

\newcommand{\mbnormal}{\texttt{mb2016}}
\newcommand{\mbrescaled}{\texttt{mb2016-rescaled}}
\newcommand{\lengthscale}{\lambda}

\DeclareSIUnit\year{yr}
\DeclareSIUnit\Msol{M_{\odot}}
\DeclareSIUnit\Zsol{Z_{\odot}}
\DeclareSIUnit\Lsol{L_{\odot}}
\DeclareSIUnit\Rsol{R_{\odot}}
\DeclareSIUnit\parsec{pc}
\DeclareSIUnit\lightyear{ly}
\DeclareSIUnit\arcsec{^{\prime\prime}}
\DeclareSIUnit\arcmin{\prm}
\DeclareSIUnit\magnitudes{mag}
\DeclareSIUnit\dex{dex}
\DeclareSIUnit\erg{erg}
\DeclareSIUnit\au{AU}
\DeclareSIUnit\jansky{Jy}
\DeclareSIUnit\rydberg{Ry}
\DeclareSIUnit\counts{counts}

\newcommand{\kpc}{\kilo\parsec}

\newcommand{\pervolcm}{\cm\tothe{-3}}
\newcommand{\Myr}{\mega\year}
\newcommand{\Gyr}{\giga\year}

\newcommand{\solMass}{\Msol}
\newcommand{\solLum}{\Lsol}

\newcommand{\Msolpyear}{\Msol\per\year}

\newcommand{\orcid}[1]{\href{https://orcid.org/#1}{\includesvg[width=10pt]{logos/ORCID-iD_icon_vector}}}

\title[An evolutionary model of the MW's hot CGM]{An evolutionary model of the Milky Way's hot circumgalactic medium}

\author[S.K. Sipma et al.]{
Sten K. Sipma$^{1}$\thanks{E-mail: sipma@astro.rug.nl}\orcid{0009-0003-5192-6858},
Filippo Fraternali$^{1}$\orcid{0000-0002-0447-3230},
Kate H. R. Rubin$^{2}$\orcid{0000-0001-6248-1864},
Federico Marinacci$^{3,4}$\orcid{0000-0003-3816-7028}
\\
$^{1}$Kapteyn Astronomical Institute, University of Groningen, Postbus 800, NL-9700 AV Groningen, The Netherlands\\
$^{2}$Department of Astronomy, San Diego State University, San Diego, CA 92182 USA\\
$^{3}$Department of Physics \& Astronomy ‘Augusto Righi’, University of Bologna, via Gobetti 93/2, I-40129 Bologna, Italy\\
$^{4}$INAF, Astrophysics and Space Science Observatory Bologna, via P. Gobetti 93/3, I-40129 Bologna, Italy\\
}

\date{Accepted XXX. Received YYY; in original form ZZZ}

\pubyear{2026}

\begin{document}%
\label{firstpage}
\pagerange{\pageref{firstpage}--\pageref{lastpage}}
\maketitle

\begin{abstract}
	The properties and the evolution of the volume-filling, hot \acf{CGM} around the Milky Way are poorly
	constrained, even though this is a crucial component in understanding how gas accretes
	onto the Galactic disc to form stars. In this work, we use a
	semi-analytic model that evolves a spherically symmetric gas in approximate
	hydrostatic equilibrium through radiative cooling, photoionization (both
	from Galactic sources and background) and mechanical heating from a central
	source. We apply this model to the hot \ac{CGM} of the Milky Way, where the
	density, temperature and metallicity are constrained from observations. Our results
	show that in
	the absence of mechanical feedback, gas in the inner few \si{\kpc} cools
	within a few tens of \si{Myr}. Gas beyond \SI{10}{kpc} cools slowly and flows
	towards the centre at a constant rate of $\dot M \approx 0.1$ to
	\SI{0.7}{\Msolpyear} depending on the total mass of the hot \ac{CGM}. Within our modelling framework, the continuous mechanical feedback that is needed to
	balance the cooling in the centre is $10^{41-41.75}\,\si{\erg\per\s}$. We show that, when averaged over a full duty
	cycle, the amount of feedback provided by various possible sources that created
	the Fermi bubbles, is similar to or more energetic than the feedback that our
	model requires to balance the cooling. This suggests the source that created
	the Fermi bubbles has sufficient energy to stop the cooling of the Milky Way's hot \ac{CGM}, and could
	form a self-regulating cycle with the hot \ac{CGM} that could keep it stable for several \si{\Gyr}.
\end{abstract}

\begin{keywords}
	Galaxy: halo --
	Galaxy: evolution --
	hydrodynamics --
	radiation mechanisms: thermal
\end{keywords}



\section{Introduction}\label{sec:Introduction} 
The Milky Way is expected to be surrounded by a hot ($T\sim \SI{e6}{K}$),
diffuse, volume filling gas in approximate hydrostatic equilibrium, typically
referred to as the hot-phase of the
\acf{CGM}~\citep{spitzer1956PossibleInterstellarGalactic,white1978CoreCondensationHeavy,tumlinson2017CircumgalacticMedium}. This hot \ac{CGM} is
thought to have formed around redshift
$z=1.5-2$~\citep{vandevoort2011DropCosmicStar,gronnow2024DensityMilkyWays}
through hot-mode accretion; the accretion of gas from the
\ac{IGM} into the dark matter halo and subsequent shock heating of this gas to
the halo's virial temperature. The gas then cools slowly, with cooling times
greater than the Hubble time, and is therefore still present
today~\citep{birnboim2003VirialShocksGalactic,dekel2006GalaxyBimodalityDue,keres2009GalaxiesSimulatedLCDM,correa2018FormationHotGaseous}.

Hot gas at such temperatures will be observable in the soft X-ray
($0.5-\SI{2.0}{\keV}$), but observing the full extent of the hot \ac{CGM}
around the Milky Way has been proven difficult due to extremely low densities
($n < 10^{-4}\ \si{\pervolcm}$) beyond
\SI{100}{\kpc}~\citep{gatto2013UnveilingCoronaMilky,miller2015ConstrainingMilkyWays}.
It is still possible to constrain the gas density profile using
line emission~\citep{miller2015ConstrainingMilkyWays,li2017PropertiesGalacticHot,locatelli2024WarmhotCircumgalacticMedium}
or absorption in background quasar
sightlines~\citep{gupta2012HugeReservoirIonized,miller2013StructureMilkyWays}.
Both techniques are sensitive to the highest density gas along the line of
sight, and since the density of the \ac{CGM} is highest close to the Milky Way,
these measurements are typically biased towards the densest gas within a few
tens of \si{kpc} from the Sun~\citep{locatelli2024WarmhotCircumgalacticMedium}.

The temperature of the Milky Way's hot \ac{CGM} has been relatively well constrained to a
value of
$T=10^{6.3}\si{\kelvin}$~\citep{henley2013XMMNEWTONSURVEYSOFT,ponti2023AbundanceTemperatureOuter},
but metallicity estimates still have some uncertainties. Most recent measurements,
using data from eROSITA~\citep{ponti2023AbundanceTemperatureOuter}, indicate
that the metallicity of the hot \ac{CGM} should be between
$Z={0.05}-\SI{0.1}{\Zsol}$. This value is considerably lower compared to the
metallicity that has typically been assumed to constrain the density of the
\ac{CGM}~\citep[for example $Z=\SI{0.3}{\Zsol}$;][]{gupta2012HugeReservoirIonized,miller2013StructureMilkyWays,miller2015ConstrainingMilkyWays,li2017PropertiesGalacticHot}.

Despite the difficulty in observing the hot \ac{CGM} of the Milky Way, studying
this component can give us insights into several open problems in the field of
galaxy formation and evolution. Estimates of the total baryonic mass within the
Milky Way \ac{CGM} suggest that it can be a significant contributor to the total
baryonic mass budget of our
galaxy~\citep{miller2013StructureMilkyWays,li2017PropertiesGalacticHot,locatelli2024WarmhotCircumgalacticMedium,zhang2024HotCircumgalacticMediuma}.
Constraining the mass therefore can help solve the `missing' baryon problem,
where the fraction of observed baryonic to dark matter in galaxies is
significantly below the expected cosmic baryon fraction~\citep[ratio of baryonic to total density $f_{\rm b}  = \Omega_{\rm b} / (\Omega_{\rm b}+\Omega_{\rm dm}) \approx 0.157$,][]{planckcollaboration2020Planck2018Results}
predicted by
cosmological
models~\citep{mcgaugh2009BARYONCONTENTCOSMIC,dai2010BARYONFRACTIONSCLUSTERS}.
Since the hot \ac{CGM} is volume filling, most of the mass resides at -- and it is
possibly beyond -- the virial radius ($\sim 280\, {\rm kpc}$), where observations have the most difficulty to
constrain the density~\citep{locatelli2024WarmhotCircumgalacticMedium}. An
alternative approach is to observe other galaxies with halo masses similar to
the Milky Way, and estimate the baryonic mass around these galaxies.

Until recently, the only observational evidence of a hot \ac{CGM} around other
spiral galaxies was based on massive ($M_\star > \SI{2e11}{\Msol}$) low
redshift spirals, and only up to a fraction of their virial
radii~\citep[e.g.][]{dai2012XMMNewtonDetectsHot,bogdan2013HotXRayCoronae,li2016CircumGalacticMediumMASsive,anderson2016DeepXMMNewtonStudy}.
However, using XMM-Newton, \citet{das2020WarmHotExtendedMassive} were able to
detect hot gas up to $r\sim\SI{150}{kpc}$ around NGC 3211. Even more recently,
by stacking the X-ray emission around tens of thousands Milky Way-mass galaxies
observed by eROSITA, \citet{zhang2024HotCircumgalacticMediuma} detected a
signal all the way up to the virial radius, providing the first observational
evidence that the hot \ac{CGM} around Milky Way-mass galaxies extends
throughout the entire dark matter halo. The hot gas mass within the virial
radius of these stacked atmospheres (assuming a metallicity of
$Z=\SI{0.3}{\Zsol}$, but lower values would result in even higher masses) is
estimated to be of the same order as the stellar mass, and if the density
profiles are extrapolated to roughly three times the virial radii, the total
baryonic mass would reach the cosmic baryon fraction.

Another technique that can be employed to observe the hot \ac{CGM} around other
galaxies, is through the \ac{tSZ}
effect~\citep{sunyaev1970SpectrumPrimordialRadiation}. Since the signal of this
effect scales with gas mass and temperature, it is most easily detected in
massive haloes, like clusters \citep{vanderlinde2010GalaxyClustersSelected}, or
massive early-types \citep{greco2015STACKEDTHERMALSUNYAEV}. In recent
years, these limits are being pushed closer to Milky Way-mass galaxies.
\citet{bregman2022HotExtendedGalaxy} stacked 11 local $L^\star$-galaxies
showing that the total mass in the hot phase within \SI{250}{\kpc} is roughly
\SI{30}{\percent} of the cosmological baryon
fraction. This mass is about equal to the mass in stars, disc gas and warm halo
gas combined and the remaining mass that is needed to reach the cosmic baryon
fraction is likely hot and extends up to $400-\SI{500}{\kpc}$. A larger sample of
\ac{tSZ} galaxies up to $z<0.5$ was assembled
by~\citet{das2023DetectionThermalSunyaevZeldovich}, detecting a signal for
stacked galaxies with stellar mass $M_\star > 10^{10.6} M_\odot$ (and upper
limits for lower masses), and~\citet{das2025ThermalSunyaevZeldovichEffect},
pushing this further to a detection in the $M_\star =
	10^{10.6\pm0.2}\si{\Msol}$ stack. Together with the X-ray observations, these
constraints give valuable insight into the total mass that is contained in the
hot phase of the \ac{CGM}. It is also worth noting that, while the above works
mostly focus on the hot phase, the cool ($\SI{e4}{}\leq T \leq \SI{e5}{}$) phase can also contribute significantly
to the total baryonic mass around
galaxies~\citep[e.g.][]{wakker2009RelationshipIntergalacticVI,stocke2013CharacterizingCircumgalacticMedium,werk2014COSHalosSurveyPhysical,prochaska2017COSHalosSurveyMetallicities,lehner2020ProjectAMIGACircumgalactic}.

In addition to the missing baryon problem, we can also study the interaction
between the hot \ac{CGM} and the Milky Way disc, and learn how low-metallicity gas accretes
onto the Milky Way to sustain the star
formation~\citep{fraternali2012EstimatingGasAccretion} and reproduce the
chemical distribution of stars in the
disc~\citep{schonrich2009ChemicalEvolutionRadial,kubryk2013RadialMigrationBardominated,schonrich2017UnderstandingInverseMetallicity}.
Some proposed mechanisms for gas accretion are fountain
accretion~\citep{marinacci2010ModeGasAccretion,fraternali2013IonizedAbsorbersEvidence,fraternali2017GasAccretionCondensation},
cooling
flows~\citep{stern2019CoolingFlowSolutions,hafen2022HotmodeAccretionPhysics}
and
precipitation~\citep{sormani2019EffectRotationThermal,voit2021GraphicalInterpretationCircumgalactic}.
All these scenarios require us to understand the properties of the gas in the
hot \ac{CGM}, as it is considered to be the primary supply of `fresh' gas for
star formation in the Milky Way.

Finally, studies of the \ac{CGM} can also help us constrain galactic feedback through the
impact that it has on the gas in the \ac{CGM}. A primary example of such
feedback events are the Fermi and eROSITA
bubbles~\citep{su2010GIANTGAMMARAYBUBBLES,predehl2020DetectionLargescaleXray},
large $\gamma$- and X-ray emitting bubbles of hot gas that can be found
reaching up to \SI{10}{kpc} into the \ac{CGM} above and below the galactic
centre. The exact origin of these bubbles remains uncertain. The
primary candidates are past energetic episodes inside the galactic centre,
either due to nuclear star
formation~\citep[e.g.][]{crocker2015UnifiedModelFermi} or to an \ac{AGN} phase
of~\sgrAs~\citep[for
	example][]{zhang2020SimulatingFermiBubbles,yang2022FermiEROSITABubbles}. To
properly estimate the energy contained in the bubbles, and understand the
mechanism that created them, we need to fully characterise the \ac{CGM} into
which they expand.

Several models of the hot \ac{CGM} have been developed to describe its physical
properties as a function of radius. Common assumptions for these models are
spherical symmetry and hydrostatic equilibrium (or steady state), with
variations that are primarily set by the additional assumptions made. For
instance, the family of models
from~\citet{ostriker2005SimpleAccurateModel,shaw2010ImpactClusterPhysics,flender2017ConstraintsOpticalDepth}
assume a polytropic equation of state, whereas other models adopt an isothermal
\citep{faerman2017MassiveWarmHot} or isentropic
\citep{faerman2020MassiveWarmHot} \ac{CGM}.
Models with the entropy profile of the gas set by the precipitation
limit~\citep{sharma2012ThermalInstabilityFeedback,voit2019AmbientColumnDensities}
have also been proposed, together with models assuming a steady-state inflow of
material from the \ac{IGM} towards the central galaxy -- a so-called cooling
flow~\citep{stern2019CoolingFlowSolutions} -- instead of hydrostatic
equilibrium. We refer the reader to~\citet{singh2024ComparisonModelsWarmhot} for
a comparison of the gas properties and predicted observables from this array of
models.

\citet{binney2009HighvelocityCloudsForm} perform a linear perturbation analysis on a
steady-state hot \ac{CGM} with an isothermal and adiabatic equation of state.
This analysis shows thermal instabilities around the Milky Way can only form in
an adiabatic \ac{CGM}. In such a system, however, cold clouds with sizes similar
to high-velocity clouds can only form at radii larger than \SI{100}{\kpc}.
Given that all known high-velocity clouds are within approximately \SI{10}{\kpc}
from the Milky Way's disc
\citep{wakker2007DistancesGalacticHighVelocity,putman2012GaseousGalaxyHalos},
\citet{binney2009HighvelocityCloudsForm} conclude that it is unlikely the
thermal instabilities can grow in these types of systems~\citep[but also see][]{voit2021GraphicalInterpretationCircumgalactic}.

The different model from \citet{lochhaas2018FastWindsDrive} assumes the hot
\ac{CGM} is entirely a hot expanding
wind~\citep{chevalier1985WindStarburstGalaxy} powered by energy from Galactic
supernovae. Such a wind would then slowly grow to hundreds of \si{kpc} over
$5-\SI{10}{\Gyr}$, and is able to account for many of the observed \ac{CGM}
properties.

A more general time-dependent model that describes the evolution of a hot
spherically symmetric gas is formulated
by~\citet{kaiser2003SimpleModelsCooling}. They assume the gas to be in
(quasi-)hydrostatic equilibrium, where it slowly loses entropy by radiative
cooling. This model has been used to study the evolution of the \ac{ICM} in the
Hydra galaxy cluster to predict when the next \ac{AGN} outburst would occur.
This model was also used to demonstrate how the entropy profile of an initially
adiabatic, or nearly adiabatic, system changes due to radiative
cooling~\citep{binney2009HighvelocityCloudsForm}.

In this work, we use the evolutionary model presented by
\citet{kaiser2003SimpleModelsCooling} to study how the Milky Way's hot \ac{CGM}
evolves. To answer this question, we apply the model to an initial density
distribution motivated by
observations~\citep{miller2016INTERACTIONFERMIBUBBLES,zhang2024HotCircumgalacticMediuma}.
In addition to evolution in isolation, we also study how feedback from the Milky Way
impacts the evolution that follows from our model. We therefore extend the model from
\citet{kaiser2003SimpleModelsCooling}, to include heating from photoionization, both from the extragalactic \ac{UVB}~\citep{haardt2012RadiativeTransferClumpy}
and Galactic
sources~\citep{cervino2002EvolutionarySynthesisModels,kannan2016GalaxyFormationLocal},
and mechanical heating from a source in the centre of the system. We can then
apply this model to the same initial conditions,
and investigate the key differences with the model not including feedback.

The paper is organised as follows. In Section~\ref{sec:Methods} we present the model, the details of the heating
and cooling processes and our initial conditions based on Milky Way
observations. The results and implications of this model for the Milky Way are
shown in Section~\ref{sec:Results}. In Section~\ref{sec:Discussion}, we
demonstrate the differences with other initial conditions, discuss our
implementation of mechanical feedback and briefly discuss various
assumptions of our model. Our conclusions will be presented in
Section~\ref{sec:Conclusions}.


\section{Methods}\label{sec:Methods} 
In this section, we present an analytic model for the evolution of a hot
\ac{CGM}, which is an extension of the model by
\citet{kaiser2003SimpleModelsCooling}. Then, we describe the various cooling
and heating processes that govern the evolution of the model. Finally, we
choose initial conditions based on observations, such that we can apply our
model to the hot \ac{CGM} of the Milky Way.

\subsection{The evolutionary model}\label{sec:the-model}%
The model that we use to describe the evolution of the hot \ac{CGM} was
introduced by~\citet{kaiser2003SimpleModelsCooling}, and in this work, we apply
it to the hot \ac{CGM} of the Milky Way, based on recent observational
constraints~\citep{miller2015ConstrainingMilkyWays,zhang2024HotCircumgalacticMediuma}.
In addition, we extend this model by implementing heating of the gas due to
photoionizing radiation coming from the extragalactic \ac{UVB} and Galactic sources, and
mechanical heating coming from a central source.

\subsubsection{Hydrostatic CGM}\label{sec:static-cgm}%
As described by \citet{kaiser2003SimpleModelsCooling}, the model assumes a
spherically symmetric, non-rotating gas atmosphere that at each point in time is in
hydrostatic equilibrium with a static, gravitational potential, which we
assume to be dominated by the dark matter halo of the galaxy 
$\Phi_{\rm DM}(r)$. For such a gas, the equation for hydrostatic equilibrium
and the continuity equation are given by
\begin{equation}
	\dv{P(r)}{r} = -\rho(r)\dv{\Phi_{\rm DM}(r)}{r},
	\label{eq:hydrostatic-equilibrium}
\end{equation}
and
\begin{equation}%
	\dv{M(r)}{r} = 4\pi r^2\rho(r),
	\label{eq:mass-shells}
\end{equation}
where, $P(r)$ and $\rho(r)$ are the pressure and density of the gas at radius
$r$, and $M(r)$ the total gas mass inside $r$. We use the entropy index
$\sigma \equiv P \rho^{-\gamma}$, where $\gamma=5/3$ is the adiabatic index of
a monatomic gas, to rewrite Equations~\ref{eq:hydrostatic-equilibrium}
and~\ref{eq:mass-shells} into a form which does not explicitly depend on the
gas density
\begin{equation}
	\dv{P}{r} = -\pqty{\frac{P}{\sigma}}^{1/\gamma}\dv{\Phi_{\rm DM}}{r},
	\label{eq:hydrostatic-equilibrium-entropy}
\end{equation}
and
\begin{equation}%
	\dv{M}{r} = 4\pi r^2\pqty{\frac{P}{\sigma}}^{1/\gamma},
	\label{eq:mass-shells-entropy}
\end{equation}
where we also drop, for ease of notation, the explicit dependence on $r$. Solving these coupled
differential equations will give us $P(r)$ and $M(r)$, given initial
conditions for $P(0)$ and $M(0)$ and some known relation between the
entropy index and gas mass $\sigma(M)$ that we specify in
Section~\ref{sec:time-evolution}. At $r=0$  we impose $M(0) \equiv
	0$, however we do not have a similar condition for the pressure at this
boundary. To find the value of $P(0)$ we proceed by assuming that the \ac{CGM}
has a fixed total mass $M_{\rm CGM}$ at some outer radius $R_{\rm out}$, such
that $M(R_{\rm out}) \equiv M_{\rm CGM}$. At this radius, we assume the gas to
be in pressure equilibrium with the \ac{IGM} $P(R_{\rm out}) \equiv P_\infty$,
where $P_\infty$ is the pressure of the gas in the \ac{IGM}. We derive the values of
$M_{\rm CGM}$, $R_{\rm out}$ and $P_\infty$ on observations of
the Milky Way (see
Section~\ref{sec:initial-conditions-for-the-milky-way} for details).

The conditions $M(0)$ and $P(R_{\rm out})$ specify a
\textit{two-point boundary value problem}, that we solve using a
\textit{shooting method}, which iteratively guesses $P(0)$ until the outer
boundary condition is satisfied~\citep{press2007NumericalRecipesArt}. Because
$P_\infty$ is multiple orders of magnitude lower than 
$P(0)$, we require a high order integrator to solve the system without being
dominated by the uncertainty in the integrator\footnote{For this work, we use
	an eighth-order Dormand-Prince~\citep{dormand1980FamilyEmbeddedRungeKutta}
	embedded method, described in Section~17.2.4
	of~\citet{press2007NumericalRecipesArt}.}.

\begin{figure}
	\begin{center}
		\begin{tikzpicture}[ align=center, minimum width=3.3cm, minimum height=2cm ]
			\matrix[row sep=1.5cm, column sep=1.0cm, nodes={anchor=center}] {
				\node (start) [startstop] {\textbf{Start} evolution with \\ $\sigma_0(M), M_{\rm CGM} , P_\infty$\\(Section~\ref{sec:initial-conditions-for-the-milky-way})}; &                                                              \\
			\node (integrate) [process] {\textbf{Integrate}  $\sigma_t(M) $, \\ to get $P_t(0), P_t(r)$ \\and $ M_t(r)$\\(Equations~\ref{eq:hydrostatic-equilibrium-entropy},~\ref{eq:mass-shells-entropy})};             &
			\node (compute) [process] {\textbf{Compute} $\mathbb{L}(r)$ (Section~\ref{sec:cooling-and-heating-processes}),\\$t_\epsilon(r)$ (Equation~\ref{eq:characteristic-timescale}) and \\$ \delta t$ (Section~\ref{sec:time-evolution})};
            \\
			\node (update) [process] {\textbf{Compute}  $\sigma_{t+\delta t}(M)$\\(Equation~\ref{eq:entropy-update-numerical})};                                                             &
			\node (check) [decision] {\textbf{Check} \\$\delta t < \delta t_{\rm stop}$ or \\$N_{\rm iter} > N_{\rm iter,max}$ };                            \\
			& \node (stop) [startstop] {\textbf{Stop} evolution}; \\
			};
			\node[draw, dashed, fit=(integrate) (compute) (update) (check), inner xsep=1em, inner ysep=1em] {};
			\draw [arrow] (start) -- (integrate);
			\draw [arrow] (integrate) -- (compute);
			\draw [arrow] (compute) -- (check);
			\draw [arrow] (check) -- node[condtext, anchor=east] {yes} (stop);
			\draw [arrow] (check) -- node[condtext, anchor=south] {no} (update);
			\draw [arrow] (update) -- node[condtext, anchor=west,align=left] {$t \leftarrow t+\delta t$\\$N_{\rm iter} \leftarrow N_{\rm iter} + 1$} (integrate);
			\draw [arrow] (start) -- node[condtext, anchor=west,align=left] {$t = 0$\\$N_{\rm iter}=0 $} (integrate);
		\end{tikzpicture}

	\end{center}
	\caption{
		A schematic overview of the global steps of our evolutionary model, described
		in Section~\ref{sec:the-model}. The evolution starts with initial conditions
		described in Section~\ref{sec:initial-conditions-for-the-milky-way},
		integrates the entropy-index to obtain pressure and mass at time $t$ according
		to Section~\ref{sec:static-cgm}. The physical processes that are included in
		the computation of $\mathbb{L}(r, \ldots)$ are described in
        Section~\ref{sec:cooling-and-heating-processes}. Finally, the steps
        related to the time evolution, including the stopping conditions, are
        described in Section~\ref{sec:time-evolution}.
	}\label{fig:flowchart}
\end{figure}
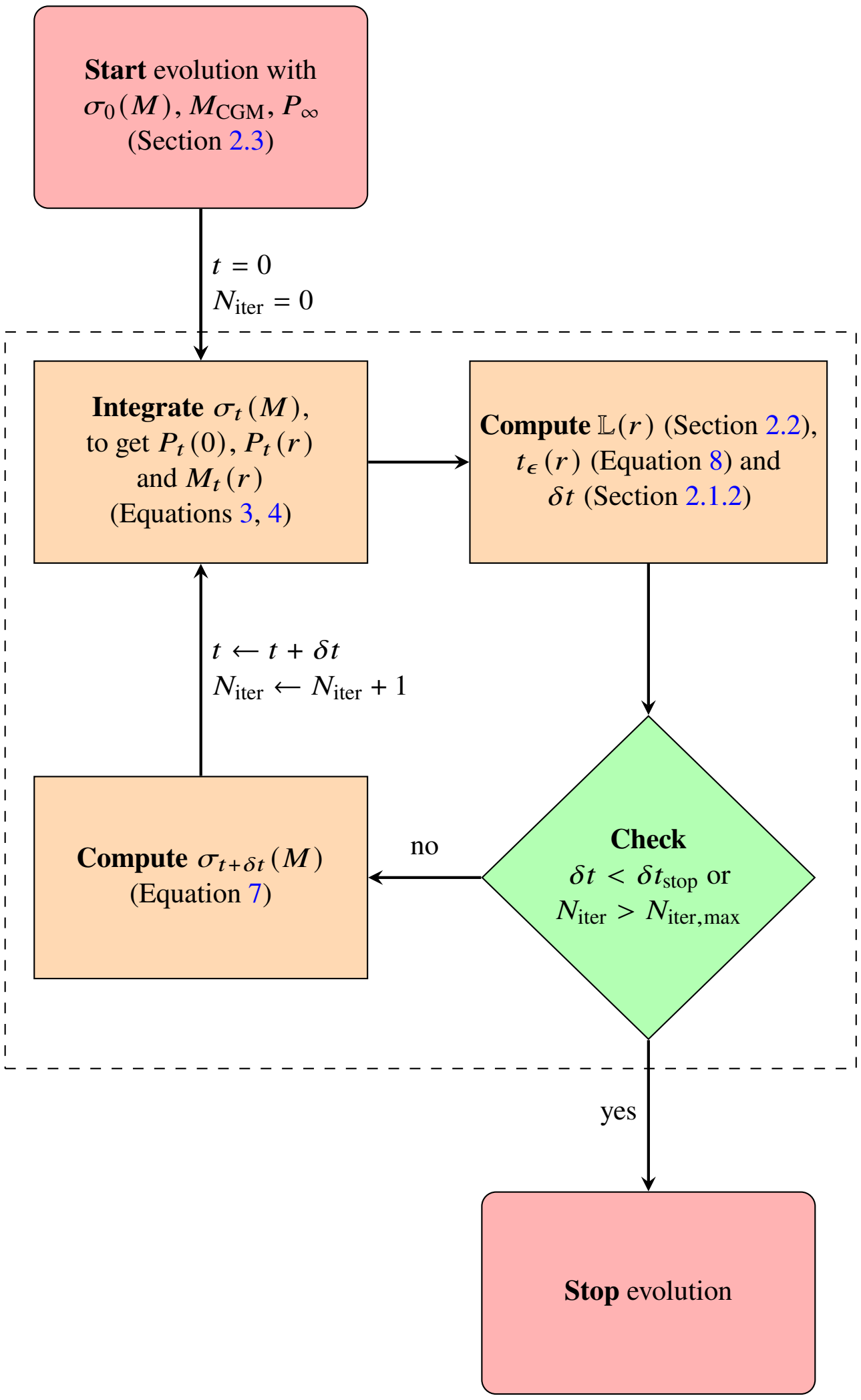

\subsubsection{Time evolution}\label{sec:time-evolution}%

After obtaining the pressure and mass using the numerical integration described
in Section~\ref{sec:static-cgm}, we calculate how the entropy of this gas
changes due to various heating\footnote{We note that heating
refers to external heating processes like mechanical heating or photoionization,
not compressional heating that is implicit in the equations for hydrodynamics.}
and cooling processes. The change in specific
entropy $\mathcal{S} = C_{\rm V}\ln\sigma + \text{constant}$ (where $C_{\rm
			V}$ is the specific heat under constant volume) is given by the heat equation
\citep{clarke2007PrinciplesAstrophysicalFluid}
\begin{equation}%
    \rho T \dv{\mathcal{S}}{t} = -\mathbb{L}\pqty{T, n_{\rm H}, Z, r, \ldots},
	\label{eq:heat-equation}
\end{equation}
where we neglect all non-radiative terms and $T$ is the temperature of the gas.
The function $\mathbb{L}$ describes the amount of energy lost ($\mathbb{L} >
0$) or gained ($\mathbb{L} < 0$) per unit time and unit volume. This function
typically depends on the internal properties of the gas, like temperature $T$,
hydrogen number density $n_{\rm H}$ and
metallicity $Z$, but in principle can also depend on external factors like 
position or the strength of an incoming radiation field. When we neglect
heating, and assume the gas is in \ac{CIE}, $\mathbb{L}$ can be written
in terms of a classical \ac{CIE} cooling function \citep[for example from
][]{sutherland1993CoolingFunctionsLowDensity} $\mathbb{L}
	\approx n_{\rm e} n_{\rm t}  \Lambda_{\rm CIE}(T, Z)$, where $n_{\rm e} $ and
$n_{\rm t} $ are the electron and atom number densities of the gas, and
$\Lambda_{\rm CIE}(T, Z)$ is the \ac{CIE} cooling function depending solely on
the temperature and metallicity of the gas.

The change in entropy-index as a function of the gas properties and
$\mathbb{L}$ can be found by substituting the time derivative of the specific
entropy in Equation~\ref{eq:heat-equation}, with $\dot{\mathcal{S}} = C_{\rm
			V}\dot{\sigma} /\sigma$ to obtain
\begin{equation}%
	\dot{\sigma} = -\sigma\frac{\gamma - 1}{k_{\rm B}} \frac{\mathbb{L}}{nT},
	\label{eq:entropy-change}
\end{equation}
where we have used $C_{\rm V} \rho = nk_{\rm B}/(\gamma-1)$, where $k_{\rm B}$
is the Boltzmann constant and $n$ is the number density of all particles in the
gas. Numerically approximating the time derivative in the above equation gives
an explicit scheme that we use to calculate how entropy changes in a time-step
of $\delta t$:
\begin{equation}%
	\sigma_{t + \delta t}(M) = \sigma_t(M) - \sigma_t(M)\frac{\gamma-1}{k_{\rm B} } \frac{\mathbb{L}}{nT} \delta t.
	\label{eq:entropy-update-numerical}
\end{equation}
Here, $\sigma_t(M)$ and $\sigma_{t + \delta t}(M)$ are the entropy-index at
time $t$ and $t+\delta t$ evaluated at a mass $M$, and all the remaining
quantities are evaluated at the time $t$ and the radius $r$ that encloses mass
$M$. We compute the number density and temperature from the pressure and
entropy-index, obtained from the integration described in
Section~\ref{sec:static-cgm}. To compute these quantities, we assume the gas is
ideal and fully ionized\footnote{In practice the gas never reaches temperatures
below $\SI{e5}{\kelvin}$, so it is always ionized.} with mean molecular weight
$\mu = 0.62$. The details on how $\mathbb{L}$ is computed are described in
Section~\ref{sec:cooling-and-heating-processes}.

With $\sigma_{t+\delta t}$ known, it is then possible to obtain the pressure
$P(r)$ and mass $M(r)$ at time $t+\delta t$ by integration of Equations~\ref{eq:hydrostatic-equilibrium-entropy} and~\ref{eq:mass-shells-entropy} at this new time. This allows us to
update the entropy-index for the next time-step, and so on. A schematic overview
of this iterative procedure can be seen in Figure~\ref{fig:flowchart}.

The right-hand side of Equation~\ref{eq:entropy-update-numerical} can be
rewritten in terms of a characteristic time-scale in which the energy or
entropy of the system changes significantly. This time-scale can be derived by
considering an almost static fluid (i.e., velocity much smaller than the sound
speed, $v \ll c_s$, which is satisfied for a quasi-hydrostatic gas distribution), and by assuming the change in internal energy (per unit
volume) is only due to radiative cooling or heating processes ($\dot \epsilon =
\mathbb{L}$). Then the ratio of the internal energy and its absolute change in
time gives a characteristic time-scale
\begin{equation}
	t_\epsilon = \frac{\epsilon}{\abs{\dot \epsilon}} = \frac{k_{\rm B} nT}{(\gamma-1)\abs{\mathbb{L}}},
	\label{eq:characteristic-timescale}
\end{equation}
where we take the absolute value of $\mathbb{L}$ since it can be positive (net cooling) or
negative (net heating). Substituting this back into
Equation~\ref{eq:entropy-update-numerical} and accounting for the sign of
$\mathbb{L}$ gives
\begin{equation}%
	\sigma_{t + \delta t}(M) = \sigma_t(M)\pqty{1 - \texttt{sign}(\mathbb{L})\frac{\delta t}{t_\epsilon}}.
	\label{eq:entropy-update-timescale}
\end{equation}
Given that it would be unphysical to have $\sigma < 0$, it is clear from this
equation that we must keep $\delta t / t_\epsilon < 1$. We can therefore use
the characteristic time-scale $t_\epsilon$ to maintain a step-size $\delta t$
which is sufficiently small. In practice, we set an upper limit $\delta t_0$, and
whenever $t_\epsilon$ becomes small enough to be of the same order of $\delta
	t_0$ (indicating that energy changes rapidly) we set $\delta t = t_\epsilon /
	10$.

We implement two conditions that control when the time evolution of the \ac{CGM}
is halted. The first is when a maximum number of iterations ($N_{\rm max}$) is
reached. The second is when the step-size $\delta t$ becomes smaller than a
fixed lower limit $\delta t_{\rm stop} = \SI{0.01}{\Myr}$\footnote{We ran tests
with lower limits that are a factor 10 and 100 below $\delta t_{\rm stop}$ to
ensure the value of the lower limit does not affect our results.}. This will
typically be the case when $\abs{\mathbb{L}}$ is very large, which means the
system is cooling or heating very rapidly. In both cases, the cooling (or
heating) is only further amplified, meaning the system will be in a runaway
situation. The state of the system in runaway cannot be described by our model
since the assumption of hydrostatic equilibrium is no longer valid. We therefore
stop the evolution, and restrict our analysis to the evolution of the hot gas
prior to the runaway state. We note that $t_{\rm end}$ in this
latter case corresponds to the time at which the hydrostatic treatment
ceases to be applicable, while $t_{\rm end}$ when the maximum number of
iterations is reached does not have an explicit physical meaning.

\subsubsection{Negative entropy gradients}\label{sec:negative-entropy-gradients}%

When the entropy changes, as a result of heating, it will become possible for
the entropy to decrease with radius $\dv*{\mathcal{S}}{r} < 0$ at which point
such a region becomes convectively
unstable~\citep{balbus1995ThermalInstability}. To avoid such instabilities, we
implement a simple method that flattens these unstable regions in analogy to
what is done by~\citet{binney2009HighvelocityCloudsForm}. At each time-step, we
identify regions with negative entropy gradients, and set the entropy-index at
each point in the region equal to $\bar\sigma$. This entropy-index corresponds to the
mass-weighted entropy of the region ($\bar{\mathcal{S}}=C_{\rm V}\ln\bar\sigma
	+ {\rm constant}$) and is given by
\begin{equation}
	\ln \bar\sigma = \frac{ \sum_i \ln[\sigma(r_i)] T(r_i) m(r_i) }{\sum_i T(r_i)m(r_i)},
	\label{eq:convection-average}
\end{equation}
where the sum is over all radii $r_i$ that are in a continuous region of
decreasing entropy and $T(r_i)$ is the temperature, $\sigma(r_i)$ the entropy
index and $m(r_i)$ the gas mass in a shell, at radius $r_i$. We compute
$m(r_i)$ from the difference in total mass between the two radii $m(r_i) \equiv
	M(r_i) - M(r_{i-1})$, with $m(0) \equiv 0$. In Section~\ref{sec:convection}, we
show how this procedure affects our results.

\subsection{Cooling and heating processes}\label{sec:cooling-and-heating-processes}%

As already mentioned, the evolution of the hot \ac{CGM} will almost exclusively be determined by the
processes that change the entropy in the system
(Equation~\ref{eq:heat-equation}). We take these processes into account using
$\mathbb{L}$, which gives the amount of energy lost ($\mathbb{L} > 0$) or
gained ($\mathbb{L} < 0$) by the system per unit volume per unit time. To
compute $\mathbb{L}$ we consider three types of processes:
\begin{enumerate}
	\item radiative cooling of the \ac{CGM} gas,
	\item heating of the gas through photoionizing radiation coming from
	      local Galactic sources and the uniform extragalactic \acf{UVB}, and
	\item mechanical heating of the gas by a central source.
\end{enumerate}
We describe processes (i) and (ii) in Subsections~\ref{sec:photoionization} to~\ref{sec:flux-calculation}, and describe (iii) in Subsection~\ref{sec:mechanical-heating}.

\subsubsection{Radiative cooling and photoionization}\label{sec:photoionization}%

\begin{figure*}
	\begin{center}
		\includegraphics{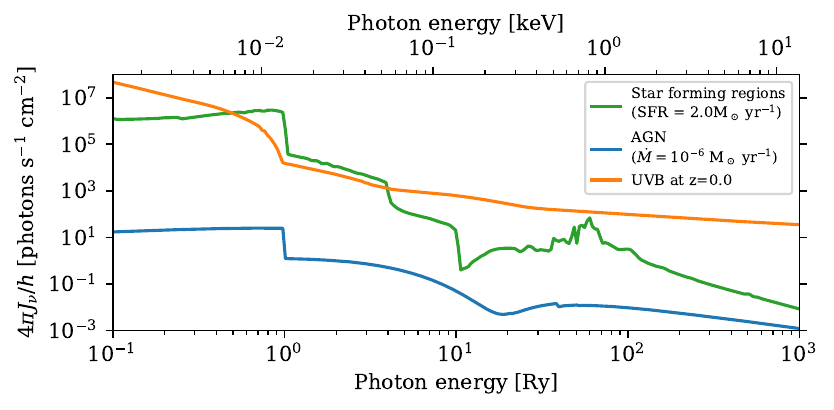}
	\end{center}
	\caption{
		Spectral flux of the photoionizing radiation emitted by a star forming
		region (green) and an \ac{AGN} (blue) at a distance of \SI{100}{kpc}
		from the source where the radiation is emitted. The photon energy
		is displayed in units of \si{\rydberg} $(= \SI{13.6}{eV})$ on the bottom
		X-axis, and in \si{keV} on top. The star forming region
		is generated with ${\rm SFR} = \SI{2}{\Msolpyear}$ and the \ac{AGN}
		has mass accretion $\dot M = \SI{e-6}{\Msolpyear}$. Both spectra
		include absorption due to gas in the Milky Way disc (see
		Section~\ref{sec:absorption-in-the-ism}). The orange line shows the
		\ac{UVB} from \citet{haardt2012RadiativeTransferClumpy} at $z=0$, which
		is constant for all radii.
	}\label{fig:photoionizing-spectrum}
\end{figure*}

We consider photoionization of the gas originating from various sources. First,
we include photoionization from the uniform  extragalactic \citet{haardt2012RadiativeTransferClumpy} \ac{UVB} at redshift $z=0$. Second, we include photoionization due to emission from
star forming regions in the Galactic disc
(Section~\ref{sec:photoionization-from-star-forming-regions}), and emission
from an \ac{AGN} (Section~\ref{sec:photoionization-from-agn}) in the centre of
the Milky Way.

We use Cloudy~\citep[v23.1,][]{chatzikos20232023ReleaseCloudy} to compute
cooling tables, which give the radiative cooling ($\Lambda$) and heating
($\Gamma$) due to photoionization for a gas with given metallicity $Z$,
temperature $T$, hydrogen number density $n_{\rm H}$ and incident photoionizing
flux (number of photons per second per unit area with energy above
\SI{13.6}{\eV}) from star forming regions $\Phi_{\rm SFR}$ and the \ac{AGN},
$\Phi_{\rm AGN}$ (for details on how we compute $\Phi$ for both processes see
Section~\ref{sec:flux-calculation}). The spectra of all the photoionizing
components can be seen in Figure~\ref{fig:photoionizing-spectrum}. Finally, the
energy per unit volume per unit time that is removed or added to the system due
to radiative processes is found by subtracting the heating from the cooling:
$\mathbb{L}_{\rm rad}(Z, T, n_{\rm H}, \Phi_{\rm SFR}, \Phi_{\rm AGN}) =
	\Lambda - \Gamma$, where $\Lambda$ and $\Gamma$ also depend on the input
parameters.

We generate the cooling and heating rates on a logarithmic grid, varying $\log
(n_{\rm H}/ \si{\pervolcm})$ from $-6$ to $0$ with steps of \SI{0.5}{\dex},
$\log (T/\si{\kelvin})$ from $4$ to $8$ with steps of $\SI{0.05}{\dex}$, and $\log
(Z/\si{\Zsol})$ from $-1$ to 0 with steps of $0.5$. Both logarithmic
fluxes, $\log(\Phi/[\si{photons\,\s\tothe{-1}\cm\tothe{-2}}])$, are varied between 3 and 10 with
steps of \SI{1}{\dex}. Below the flux of the \ac{UVB} ($\Phi_{\rm UVB} \approx
\SI{e4}{photons\,\s\tothe{-1}\cm\tothe{-2}}$), our sampling does not have to be as
fine so we just add a single grid point at $\log\Phi=0$.
The next sections describe the details of the individual spectra emitted by
star forming regions and the \ac{AGN}, that are used as an input to the Cloudy
calculations.

\subsubsection{Photoionization from star forming regions}\label{sec:photoionization-from-star-forming-regions}%

High energy radiation associated with star forming regions is emitted at
several stages in the evolution of high-mass stars. For a recently formed
population of stars ($\leq \SI{1}{\Myr}$ ago), O and B type stars are the
primary sources of photoionizing UV radiation due to their high temperatures.
When these stars explode as core-collapse supernova, soft X-rays are mostly emitted during the Sedov
phase of the remnant. This radiation originates from the hot gas in the
expanding bubble and the shock front of the \ac{SNR}. Finally, the mechanical
energy that is released in the \ac{ISM} by stellar winds, supernovae and
Wolf-Rayet stars is able to heat the gas to temperatures where it will emit
soft X-rays.

\citet{cervino2002EvolutionarySynthesisModels} model the spectral evolution of
a \ac{SSP}, including O/B stars, \ac{SNR} and heating of the \ac{ISM}
(parametrized by $\epsilon^{\rm X}_{\rm eff}$, the fraction of mechanical
energy that heats the \ac{ISM}). They track the evolution of such an
\ac{SSP} up to \SI{10}{Myr} from birth, at which point the contributions of the
high-mass stars and \ac{SNR}s are negligible. After \SI{10}{\Myr} the spectrum
of the \ac{SSP} is therefore determined by the heating of the \ac{ISM} and
stays relatively constant due to long cooling times. We combine the
spectrum of these \ac{SSP}s at several evolutionary stages into one spectrum to
obtain the spectrum of a stellar population that has been forming stars at a
constant rate for \SI{10}{Myr}.

The total spectral flux $F_T(\nu)$ (as a function of frequency $\nu$) of a
stellar population of age $T$ with star formation history $\Psi(t)$ is given by
\begin{equation}
	F_T(\nu) = \int_0^T \Psi(t)f_t(\nu)\dd t,
	\label{eq:spectrum-combine}
\end{equation}
where $f_t(\nu)$ is the spectral flux (per unit mass) of a \ac{SSP} with age $t$
given by~\citet{cervino2002EvolutionarySynthesisModels}\footnote{\url{https://cab.inta-csic.es/users/mcs/SED/tables.html\#SED}}.
Following~\citet{cantalupo2010StarsQuenchingStars}, we obtain the spectral flux
for a \ac{SSP} of solar metallicity and fraction of mechanical energy that heats
the \ac{ISM} $\epsilon^{\rm X}_{\rm eff} = 0.05$. We additionally renormalise
the spectrum to follow a Chabrier
\ac{IMF}~\citep{chabrier2003GalacticStellarSubstellar}, which has an higher
number of high mass stars compared to the Salpeter
\ac{IMF}~\citep{salpeter1955LuminosityFunctionStellar} that was used to create
the \ac{SSP}s.

We note that other works that include photoionization for star forming
regions~\citep{cantalupo2010StarsQuenchingStars,kannan2014GalaxyFormationLocal}
assume that the spectrum for a \ac{SSP} of age \SI{5}{Myr} is representative
for a star-forming region. Compared to our spectrum that combines multiple
\ac{SSP}s, these spectra will have weaker emission between
$1-\SI{10}{\rydberg}$, and slightly stronger emission above \SI{10}{\rydberg}.
Moreover, if a star forming region would continuously form stars for longer
than \SI{10}{Myr}, this would result in increased emission above
\SI{10}{\rydberg} due to the additional heating of the \ac{ISM}. However, a
detailed study of these effects are beyond the scope of this work.  We assume
the \SI{10}{Myr} star forming region to be representative for any star
forming region.

The remaining compact body at the end of the evolution of a high-mass star can
also form a binary system with a companion star that consequently can transfer
mass to the surface of the compact body resulting in the emission of hard X-ray
radiation. The properties of these X-ray binaries depend on the companion star,
and in the case of a high-mass companion, such a system will be a \acf{HMXB}. A
low-mass companion, yielding a \ac{LMXB} is much more common, but their emission
is generally negligible compared to the \ac{HMXB} in a star forming
region~\citep{kannan2016GalaxyFormationLocal}.

The contribution from \ac{HMXB}s can be modeled by a power-law for photon
energies $E > \SI{0.5}{keV}$
\citep{dotani1997ASCAObservationCygnus,gierlinski1999RadiationMechanismsGeometry,kannan2016GalaxyFormationLocal}
\begin{equation}
	F_{\rm HMXB}(E) \propto E^{-2},
	\label{eq:hmxb-powerlaw}
\end{equation}
where $E$ is the photon energy and $F$ is the flux density (energy per unit time, area
and photon energy). We normalize this spectrum, using the relation between the
\ac{HMXB} luminosity and \ac{SFR} of the region, which is given by
\citet{mineo2012XrayEmissionStarforming}:
\begin{equation}
	L_{\rm HMXB} = \SI{2.61e39}{}\pqty{\frac{\rm SFR}{\SI{1}{\Msolpyear} }} \,\si{\erg\per\s},
	\label{eq:hmxb-sfr-relation}
\end{equation}
in the energy range from $0.5$ to \SI{8.0}{keV}.

The total spectrum emitted by star forming regions is then the combination of
the stellar, the \ac{SNR}, and the \ac{HMXB} spectra. The green line in
figure~\ref{fig:photoionizing-spectrum} shows this combined spectrum, including
ISM absorption, for a \ac{SFR} of \SI{2}{\Msolpyear} at a distance of
\SI{100}{kpc}. Up to \SI{10}{Ryd}, the spectrum is dominated by the O/B stars,
from \SI{10}{Ryd} to \SI{100}{Ryd}, the \ac{SNR} component dominates, and at
energies above \SI{100}{Ryd}, the \ac{HMXB} component is dominant.

\subsubsection{Photoionization from the \acs{AGN}}\label{sec:photoionization-from-agn}%
We describe the photoionizing radiation from the central black hole using the
two-component model from \citet{bland-hawthorn2013FOSSILIMPRINTPOWERFUL}. They define
the specific photon rate as
\begin{equation}
	\begin{split}
		N_{\rm AGN}(E) & = k_1 \pqty{\frac{E}{E_1}}^{-2/3}e^{-E / E_1}                                                          \\
		               & + k_2\pqty{\frac{E}{E_2}}^{-\alpha}e^{-E / E_2} \mathcal{H}(E - E_1) {\rm\ photons\ s^{-1} \ eV^{-1}},
	\end{split}
	\label{eq:agn-spectrum}
\end{equation}
where $\mathcal{H}(x) = 1$  for $x>0$ and 0 otherwise. The first component
describes the cool, optically thin Shakura-Sunyaev spectrum observed in Seyfert
galaxies and quasars~\citep{antonucci1993UnifiedModelsActive}, with $E_1 =
	\SI{30}{eV}$. The second term is the observed emission in X-rays and
$\gamma$-rays from an AGN, with $E_2 = \SI{100}{eV}$ and $\alpha =
	1.9$~\citep{dermer1997OSSEObservationsUltraluminous}. The ratio of the relative
normalization constants $k_1$ and $k_2$ is derived from the ionizing spectrum
of NGC 1068, giving $\log k_1/k_2 = 8.8663$ \citep[see Appendix C of][]{bland-hawthorn2013FOSSILIMPRINTPOWERFUL}. The total luminosity of
the black hole depends on the accretion rate $\dot M$, and the efficiency with
which the black hole converts energy into radiation $\epsilon\approx0.05$ as
\begin{equation}
	L_{\rm AGN} = \SI{7.3e11}{} \pqty{\frac{\epsilon}{0.05}} \pqty{\frac{\dot M}{\si{\Msolpyear}}}\,\si{\solLum}.
	\label{eq:agn-luminosity}
\end{equation}
The best estimate of the accretion rate of \sgrAs~is $\dot{M} = \SI{e-9}{}-\SI{e-8}{\Msolpyear}$~\citep{eventhorizontelescopecollaboration2022FirstSagittariusEventb}, so for our fiducial model we
adopt a value of $\dot{M}_{\rm AGN} = \SI{e-8}{\Msolpyear}$.

The spectrum, including \ac{ISM} absorption, of an \ac{AGN} with an accretion
rate of \SI{e-6}{\Msolpyear} is shown in
Figure~\ref{fig:photoionizing-spectrum}. We note that this accretion rate is
two orders of magnitudes higher compared to the current accretion rate of
\sgrAs, showing that this contribution is not as relevant compared to the
photoionization from star forming regions.

\subsubsection{Absorption in the ISM}\label{sec:absorption-in-the-ism}%
The radiation coming from star forming regions and \ac{AGN} will have to pass
through the \ac{ISM} before it can photoionize gas in the \ac{CGM}. We model
the fraction of photons that escape the galaxy and reach the \ac{CGM}, at
frequency $\nu$ as \citep{cantalupo2010StarsQuenchingStars}
\begin{equation}
	f_{\rm esc}(\nu) = f_{\rm esc}^{\rm LL} + ( 1 - f_{\rm esc}^{\rm LL})e^{-\tau(\nu)},
	\label{eq:escape-fraction}
\end{equation}
where $f^{\rm LL}_{\rm esc} = 0.05$ is the escape fraction at the Lyman limit
\citep[similar to the value adopted by][]{cantalupo2010StarsQuenchingStars} and $\tau(\nu)$ is the
optical depth of the absorbing medium. We calculate $\tau(\nu)$ by summing over
the contributions from individual absorbers
\begin{equation}
	\tau(\nu) = \sum_i \tau_i(\nu) = \sum_i \sigma_i(\nu)N_i,
	\label{eq:optical-depth}
\end{equation}
where $\sigma_i$ and $N_i$ are the photoionization cross section and column
density of element $i$ respectively. For this work, we consider absorption due
to neutral H, He and O with cross-sections given by
\citet{verner1995AnalyticFITSPartial} and absorption due to dust (primarily
coming from silicate grains), with cross-sections\footnote{These cross sections are
	derived from regions with ratio of extinction to reddening in the
	V-band, $R_{\rm V} = 3.1$, representative of the diffuse Galactic \ac{ISM}.} given by
\citet{draine2003ScatteringInterstellarDusta}.

Observations of the Milky Way disc by
\citet{marasco2017DistributionKinematicsAtomic} show that the column density in
the region where star formation is most active ($r \geq \SI{3}{\kpc}$) is
around $N_{\rm HI} \approx \SI{5e20}{\cm^{-2}}$. Assuming the sources of star
formation are typically in the mid-plane of the disc, the radiation will have to
pass through approximately half the disc. We therefore take $N_{\rm HI} \approx
	\SI{2e20}{\cm^{-2}}$ as the column density that radiation coming from star
forming regions has to pass through in order to reach the \ac{CGM}. Extinction
in the centre of the galaxy is considerably higher, so for the radiation coming
from the \ac{AGN}, we take $N_{\rm HI} \approx \SI{e21}{\cm^{-2}}$. The
corresponding column densities for neutral He and O can then be calculated by
assuming the gas has solar metallicity, and abundances given by
\citet{asplund2009ChemicalCompositionSun}.

For a column density of $N_{\rm HI} \approx \SI{2e20}{\cm^{-2}}$, the optical
depth of H and He is significantly above one for photons with energies between
\SI{1}{} to \SI{10}{\rydberg}, and therefore most of the radiation is absorbed
($f_{\rm esc} = f^{\rm LL}_{\rm esc} = \SI{5}{\percent}$). The optical depth of
H and He decreases for photons between \SI{10}{} to \SI{30}{\rydberg},
resulting in an escape fraction of \SI{50}{\percent} at \SI{30}{\rydberg}. For
higher energies, absorption due to dust becomes relevant and the decrease in
$f_{\rm esc}$ is more gradual. At \SI{100}{\rydberg}, more than
\SI{80}{\percent} of the radiation will escape from the disc.

\subsubsection{Photoionizing flux}\label{sec:flux-calculation}%

\begin{figure}
	\begin{center}
		\includegraphics{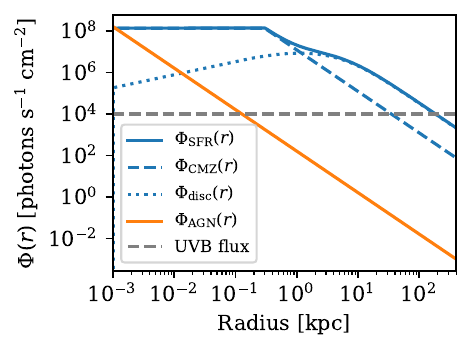}
	\end{center}
	\caption{
		Radial dependence of the photoionizing flux $\Phi$ due to star forming
		regions in the Milky Way disc and \ac{CMZ} (blue, $\Phi_{\rm SFR} =
		\Phi_{\rm CMZ} + \Phi_{\rm disc}$), and due to AGN (orange, $\Phi_{\rm
				AGN}$). The uniform \ac{UVB} from \citet{haardt2012RadiativeTransferClumpy}
		at $z=0$ ($\Phi_{\rm UVB} \approx
			\SI{e4}{photons\per\s\per\cm\tothe{2}}$) is included as the grey-dashed
		line.
	}\label{fig:local-flux}
\end{figure}

The photoionizing flux field coming from star forming regions and \ac{AGN},
as opposed to the \ac{UVB}, is not uniform and depends heavily on the distance
to the Galactic disc and the Galactic centre. We separately model the flux coming from
star forming regions and \ac{AGN}, since their emitted spectra and distribution
in the galaxy are different. For both sources, we calculate the total
photoionizing luminosity $\Phi_{0}$ that is emitted at a given \ac{SFR}  or accretion rate $\dot M_{\rm AGN}$ of
\SI{1}{\Msolpyear}. This luminosity is calculated after the absorption in
the ISM is applied. Both luminosities scale linearly with $\textrm{SFR}$ and
$\dot M_{\rm AGN}$, so therefore the photoionizing flux received at distance
$r$ from an emitting point source is given as
\begin{align}
	\Phi_{\rm SFR}(r) & = \frac{\Phi_{0,\rm SFR}}{4\pi r^2}\pqty{\frac{\rm SFR}{\SI{1}{\Msolpyear} } }\text{, and} \\
	\Phi_{\rm AGN}(r) & = \frac{\Phi_{0,\rm AGN}}{4\pi r^2}\pqty{\frac{\dot M_{\rm AGN}}{\SI{1}{\Msolpyear} } }.
\end{align}
We can model the \ac{AGN} as a point source, but have to choose a different
prescription for the more distributed star forming regions in a Galactic disc.
We consider the star forming regions to be primarily located in the
Galactic disc~\citep{bacchini2019VolumetricStarFormation} but also include a small
contribution from the \ac{CMZ} at the centre of the Milky Way. The total flux coming
from star forming regions will be the sum of the two components
\begin{equation}
	\Phi_{\rm SFR}(r) = \Phi_{\rm CMZ}(r) + \Phi_{\rm disc}(r).
\end{equation}
Both these systems are axisymmetric, so we will have to approximate their
radial dependence of the flux for our spherically symmetric model.

The size of the \ac{CMZ} is relatively small, $R_{\rm CMZ}\approx
	\SI{300}{pc}$~\citep{morris1996GalacticCenterEnvironment,henshaw2023StarFormationCentral}
compared to the total size of our system. Far away, we can
consider the \ac{CMZ} to be a point source at the origin of the system. Inside
the \ac{CMZ} we consider the flux to be constant, resulting in
\begin{equation}
	\Phi_{\rm CMZ}(r) = \frac{\Phi_{\rm 0,SFR}}{4\pi}\pqty{\frac{\rm SFR_{CMZ}}{\SI{1}{\Msolpyear} } }\begin{cases}
		\pqty{\SI{300}{pc}}^{-2} & \text{if } r \leq 300{\rm\ pc} \\
		r^{-2}                   & \text{if } r > 300{\rm\ pc}
	\end{cases},
	\label{eq:cmz-flux}
\end{equation}
where we take ${\rm SFR_{CMZ}} =
	\SI{0.07}{\Msolpyear}$~\citep{henshaw2023StarFormationCentral}.

The disc of a spiral galaxy extends for multiple \si{\kpc}, and also has a
varying \ac{SFR} density along the disc. We choose to use the flux
along the axis perpendicular to the disc ($z$), taking into account that this
is an upper limit to the flux at any other position with $R\neq0$.
In general, the normally incident flux observed at a distance $D$ from a surface
$S$ is given as
\begin{equation}
	F_\nu = \int_S I_\nu(\hat n)\, \cos\psi\ \frac{\dd A}{D^2}
\end{equation}
where $I_\nu(\hat n)$ is the specific intensity along the line-of-sight $\hat n$,
and $\psi$ the angle between $\hat n$ and the normal of the area $\dd\vb{A}$. For a razor-thin
disc with \ac{SFR} surface density at (cylindrical) radius $R^\prime$ in the
disc $\Sigma_{\rm SFR}(R^\prime)$, the specific intensity is written as
\begin{equation}
	I_\nu \equiv \frac{\mathcal{L}_\nu\Sigma_{\rm SFR}}{4\pi}
	\label{eq:specific-intensity}
\end{equation}
where $\mathcal{L}_\nu$ is the luminosity per unit \ac{SFR}. For simplicity, we
assume there is no angle dependent absorption in the disc (absorption is
already considered in the value of $\Phi_{0,\rm SFR}$), so then we can simply
integrate over all photoionizing photons ($h\nu \geq \SI{13.6}{eV}$). We obtain
the photoionizing flux by integrating over the disc in $R^\prime$ and azimuthal
angle $\phi$
\begin{equation}
	\Phi_{\rm disc}(R, z) = \Phi_{0, \rm SFR}\int_S \frac{\Sigma_{\rm SFR}(R^\prime)}{4\pi D^2}\, \cos\psi\ R^\prime\dd R^\prime \dd \phi.
\end{equation}
When we write out $\cos\psi$ and $D$ in terms of $z$ and $R^\prime$ and consider
only variation of the flux along $z$ by setting $R=0$, we get
\begin{equation}
	\Phi_{\rm disc}(z) = \frac{\Phi_{0,\rm SFR}}{2}\int_0^{R_{\rm MW}} \frac{z\,\Sigma_{\rm SFR}(R^\prime)}{\pqty{z^2 + {R^\prime}^2}^{3/2}}\, R^\prime \dd R^\prime,
	\label{eq:flux-disc-mw-final}
\end{equation}
where $R_{\rm MW}=\SI{20}{kpc}$ is the upper limit for the integration. Based on a fit to the distribution of \ac{SNR}s in the Milky Way disc~\citep{green2015ConstraintsDistributionSupernova}, we
take
\begin{equation}
\Sigma_{\rm SFR}(R) \propto \pqty{\frac{R}{R_\odot}}^\alpha\exp\pqty{-\beta\frac{R - R_\odot}{R_\odot}},
\end{equation}
where $R_\odot=8.5\si{\kpc}$ is the distance from the Sun to the Galactic centre, and the best fit is found for $\alpha=1.09$ and $\beta=3.87$. This distribution is normalized to a total \ac{SFR} of \SI{1.9}{\Msolpyear}~ \citep{chomiuk2011UnificationStarFormation}. We integrate
Equation~\ref{eq:flux-disc-mw-final} numerically for $z$ between \SI{0.02}{}
and \SI{400}{kpc} with 200 points in log space. In our model, we then
interpolate the resulting values to obtain the flux at a given
distance\footnote{At $r=0$ we extrapolate, but here $\Phi_{\rm SFR}$ is
	dominated by $\Phi_{\rm CMZ}$ so the exact value at this radius is not relevant.}.

Figure~\ref{fig:local-flux} shows each of the different components as a function
of radius, computed using the Milky Way values described in the respective
sections. We see that only at the very centre, the flux from \sgrAs~(orange) is
comparable to the flux coming from the star forming regions (blue). At all other
radii, $\Phi_{\rm SFR}$ dominates. In the centre of the system, the contribution
from the \ac{CMZ} is most important due to its proximity, but beyond a few
\si{kpc}, the disc starts to become more relevant. The photoionizing flux of the
\ac{UVB}~\citep{haardt2012RadiativeTransferClumpy} is indicated by the gray
dashed line. We note that the flux from Galactic sources is considerably higher
than the \ac{UVB} at least up \SI{100}{kpc}\footnote{The total photoionizing
flux $\Phi$ is mostly determined by photons with energy close to
\SI{1}{\rydberg}. Therefore, even though the photoionizing flux coming from
star-forming regions is higher than the flux of the \ac{UVB}, the \ac{UVB} still
dominates in the X-ray regime (\SI{10}{\rydberg} and above), as can be seen
in Figure~\ref{fig:photoionizing-spectrum}.}. This shows the critical importance
of including local sources when considering photoionization \citep[see
also][]{cantalupo2010StarsQuenchingStars}.

\subsubsection{Mechanical heating}\label{sec:mechanical-heating}%

To study the influence of heating originating from a central source in the
system on the cooling of the \ac{CGM}, we include mechanical feedback in our
model. Our implementation is motivated by \ac{AGN} feedback models from hydrodynamic cosmological simulations~\citep{springel2005ModellingFeedbackStars,weinberger2017SimulatingGalaxyFormation},
where thermal energy is injected directly into the system, weighted by a
kernel. The total power of the injecting source is given by $\dot E_{\rm
			tot}$, and we distribute this power over a 3-dimensional Gaussian kernel with
length-scale $\lengthscale$. At radius $r$, the amount of energy per second and
unit volume injected into the gas is then given by
\begin{equation}
	\dot E_\lengthscale(r) = \frac{\dot E_{\rm tot} }{\pqty{2\pi\lengthscale^2}^{3/2}}\exp\pqty{-\frac{r^2}{2\lengthscale^2}}.
	\label{eq:mechanical-energy-injection}
\end{equation}
For this work, we consider length-scales that are between $2-\SI{8}{kpc}$. At each radius, we can then find $\mathbb{L}$ by adding the injected mechanical
energy to the radiative term
\begin{equation}
	\mathbb{L} = \mathbb{L}_{\rm rad}(Z, T, n_{\rm H}, \Phi_{\rm SFR}, \Phi_{\rm AGN}) + \dot E_\lengthscale(r)
	\label{eq:final-L-computation}
\end{equation}

\subsection{Initial conditions for the Milky Way}\label{sec:initial-conditions-for-the-milky-way}%

\begin{figure}
	\begin{center}
		\includegraphics{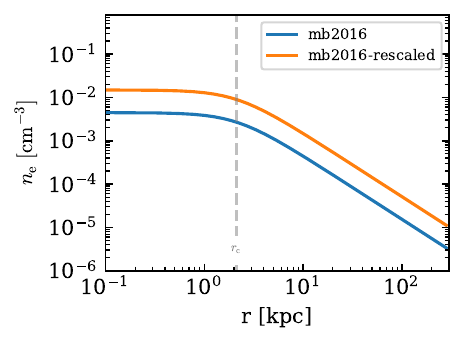}
	\end{center}
	\caption{
		$\beta$-model (Equation~\ref{eq:beta-model}) electron density profiles
		for the Milky Way hot \ac{CGM}, as derived
		by~\citet{miller2016INTERACTIONFERMIBUBBLES} (\mbnormal, blue) and a
		rescaled variant (\mbrescaled, orange). The location of the core radius
		$r_{\rm c}=\SI{2.12}{kpc}$ of the $\beta$-model is indicated by a vertical dashed
		line.
	}\label{fig:initial-density}
\end{figure}

We apply our evolutionary model described in Section~\ref{sec:the-model} to the
hot \ac{CGM} of the Milky Way. We assume the potential of the system to be static and
dark matter dominated, following a \ac{NFW} profile
\citep{navarro1997UniversalDensityProfile}. The virial mass, virial radius and
concentration of this potential are given by
\citet{bland-hawthorn2016GalaxyContextStructural}:
\begin{equation}
	M_{\rm vir} = \SI{1.3e12}{\Msol},\quad r_{\rm vir} = \SI{282}{kpc},\quad c_{\rm vir} = 10,
	\label{eq:nfw-parameters}
\end{equation}
where the virial mass also has been verified more recently using
Gaia data~\citep{posti2019MassShapeMilky}.

The gas density profile in the \ac{CGM}  is typically described by a $\beta$-model
\begin{equation}
	n(r) = n_0\bqty{1 + \pqty{\frac{r}{r_{\rm c}}}^2 }^{-3\beta/2},
	\label{eq:beta-model}
\end{equation}
where $r_{\rm c}$ is the core radius within which the density flattens out,
$\beta$ dictates the slope at $r \gg r_{\rm c}$, and $n_0$ normalizes the
distribution by specifying the density at $r=0$. The most recent
determination of the $r \gg r_{\rm c}$ regime of the \ac{CGM} of the Milky Way
is inferred based on \ac{eRASS1}
data~\citep{locatelli2024WarmhotCircumgalacticMedium}. However, the regions
close to the Galactic disc are masked out due to contaminating sources, so the
eROSITA study is not able to resolve the degeneracy between $n_0$ and $r_{\rm c}$. We
therefore adopt the values given by~\citet{miller2016INTERACTIONFERMIBUBBLES}
who simultaneously fit the hot \ac{CGM} and a Fermi bubble component using data
close to the Galactic centre, to break the degeneracy. They find $\beta
= 0.49\pm 0.01$, $r_{\rm c} = 2.12\pm
		0.22 \, {\rm kpc}$, and a central electron number density of $n_{\rm 0,e} =
	\SI{4.47e-3}{\pervolcm}$. To convert the electron number density to a total
number density $n$, we set $n = n_{\rm i} + n_{\rm e}$, and assume that the gas
contains primarily H and He and is fully ionized, yielding $n_{\rm e} \approx 1.2\,
	n_{\rm i}$.

Given the density profile described by Equation~\ref{eq:beta-model}, we can
derive the specific initial conditions required for our model: $M_{\rm CGM}$,
$P_\infty$ and $\sigma(M)$. For this, we assume that the outer radius at the
start of the evolution is equal to the virial radius of the dark matter halo $R_{\rm
			out} = r_{\rm vir}$.\footnote{We stress, that $R_{\rm out} = r_{\rm vir}$ only
	for the initial condition. At any other time-step $R_{\rm out}$ is free to
	increase or decrease, as long as the outer boundary condition on pressure is satisfied.} First, we compute the radial mass profile of the \ac{CGM} $M(r)$
by integrating over the density and obtain the total CGM mass, which is just the
mass at the virial radius $M_{\rm CGM} = M(r_{\rm vir})$. The pressure profile
is then obtained by assuming a temperature at the outer radius, $T(r_{\rm
			vir}) \equiv 10^{6.3}\
	\si{K}$~\citep{miller2013StructureMilkyWays,ponti2023AbundanceTemperatureOuter,locatelli2024WarmhotCircumgalacticMedium}
and then using the ideal gas law to find $P_\infty = P(r_{\rm vir}) = n(r_{\rm
			vir})k_{\rm B} T(r_{\rm vir})$. Integrating
Equation~\ref{eq:hydrostatic-equilibrium} from 0 to $r$ gives
\begin{equation}
	\int_{P(0)}^{P(r)} \dd P = P(r) - P(0) = - \int_0^r \rho(r^\prime)\dv{\Phi}{r^\prime}\dd r^\prime \equiv \mathcal{I}(r).
	\label{eq:ic-integral}
\end{equation}
 From this expression, we find the central pressure
by rewriting the above equation and evaluating it at $r_{\rm vir}$, giving
$P(0) = P_\infty - \mathcal{I}(r_{\rm vir})$, and then also the pressure at any radius
$P(r) = P(0) - \mathcal{I}(r)$. Finally, from $P(r)$ we can then compute the radial
entropy-index profile using $\sigma(r) \equiv P(r)\rho^{-\gamma}(r)$ and obtain
the mass-entropy relation by inverting the mass profile $\sigma(M) =
	\sigma[r(M)]$.

The total baryonic mass up to the virial radius using the density from
\citet{miller2016INTERACTIONFERMIBUBBLES} gives $M_{\rm CGM} =
	\SI{1.74e10}{\Msol}$, which is approximately \SI{10}{\percent} of the missing
baryons\footnote{We define the missing baryons as $M_{\rm b}  - M_{\rm disc}$,
	where the expected baryonic mass $M_{\rm b}$ is obtained from $M_{\rm b} =
		f_{\rm b} M_{\rm vir} / (1 - f_{\rm b})$, where $M_{\rm vir}$ is the mass
	of the dark matter halo and $f_{\rm b}$ the cosmic baryon fraction~\citep[$f_{\rm b} \approx 0.157$,][]{planckcollaboration2020Planck2018Results}. We assume the total
	observed baryonic mass in the disc to be $M_{\rm disc} \approx M_\star +
		M_{\rm gas,disc}$, where we have taken $M_\star =
		\SI{5e10}{\Msol}$~\citep{bland-hawthorn2016GalaxyContextStructural} and
	$M_{\rm gas,disc} \approx
		\SI{6e9}{\Msol}$~\citep{cimatti2019IntroductionGalaxyFormation}. } in the Milky
Way. The more recent estimates of the total baryonic mass from
\citet{zhang2024HotCircumgalacticMediuma}, based on stacking of hot atmospheres
around Milky Way-mass galaxies, are considerably higher. Depending on the
assumed metallicity and temperature of the gas they find that the hot \ac{CGM}
contributes roughly $\SI{30}{\percent}$ (for $Z=\SI{0.3}{\Zsol}$, $T = T_{\rm
			vir}$) to \SI{100}{\percent} (for $Z=\SI{0.1}{\Zsol}$, $T = 0.5T_{\rm vir}$) of
the missing mass. The actual baryonic mass of the Milky Way \ac{CGM} is likely
somewhere in between these extremes, and to study a profile that is more in
line with the~\citet{zhang2024HotCircumgalacticMediuma} values, we also consider
a rescaled variant of the \ac{CGM} of \citet{miller2016INTERACTIONFERMIBUBBLES}. This rescaled
variant has 3 times the density, and to keep roughly the total mass in metals
constant, we decrease the metallicity from $Z=\SI{0.3}{\Zsol}$ to $Z=\SI{0.1}{\Zsol}$. This metallicity
is likely also more realistic according to the most recent estimates ($Z = 0.052-\SI{0.072}{\Zsol}$)
from~\citet{ponti2023AbundanceTemperatureOuter}. We leave the structural
parameters of the distribution ($r_{\rm c}$ and $\beta$) equal to the original
values from \citet{miller2016INTERACTIONFERMIBUBBLES} (in
Section~\ref{sec:different-initial-condition} we show the results for a
larger $r_{\rm c}$){}. Figure~\ref{fig:initial-density} displays the electron
density of \citet{miller2016INTERACTIONFERMIBUBBLES} and our rescaled variant
as a function of radius.

It is likely that the present day configuration of gas in the
Milky Way's hot \ac{CGM}, is a result of a longer cooling process. In
Section~\ref{sec:different-initial-condition} we discuss the impact of an
initial condition with a core radius that is more in line with the results
from~\citet{zhang2024HotCircumgalacticMediuma} and that could also represent an
earlier stage of the Milky Way's hot \ac{CGM}.


\section{Results}\label{sec:Results} 
In this section we present the results of our semi-analytical evolutionary
model applied to the hot \ac{CGM} of the Milky Way. We first describe our
results for a \ac{CGM} that only evolves via radiative cooling (Section~\ref{sec:pure-cooling}). Then, we add
heating in the form of local sources and the \ac{UVB} (Section~\ref{sec:fiducial-model}). Finally, we explore the
impact of a heating source in the centre of the system and compare our results
with sources of feedback in the \ac{GC} of the Milky Way
(Sections~\ref{sec:heating},~\ref{sec:balancing-heating-and-cooling}
and~\ref{sec:comparison-with-fermi-bubbles}).

\subsection{Pure cooling}\label{sec:pure-cooling}%

\begin{figure*}
	\begin{center}
		\includegraphics{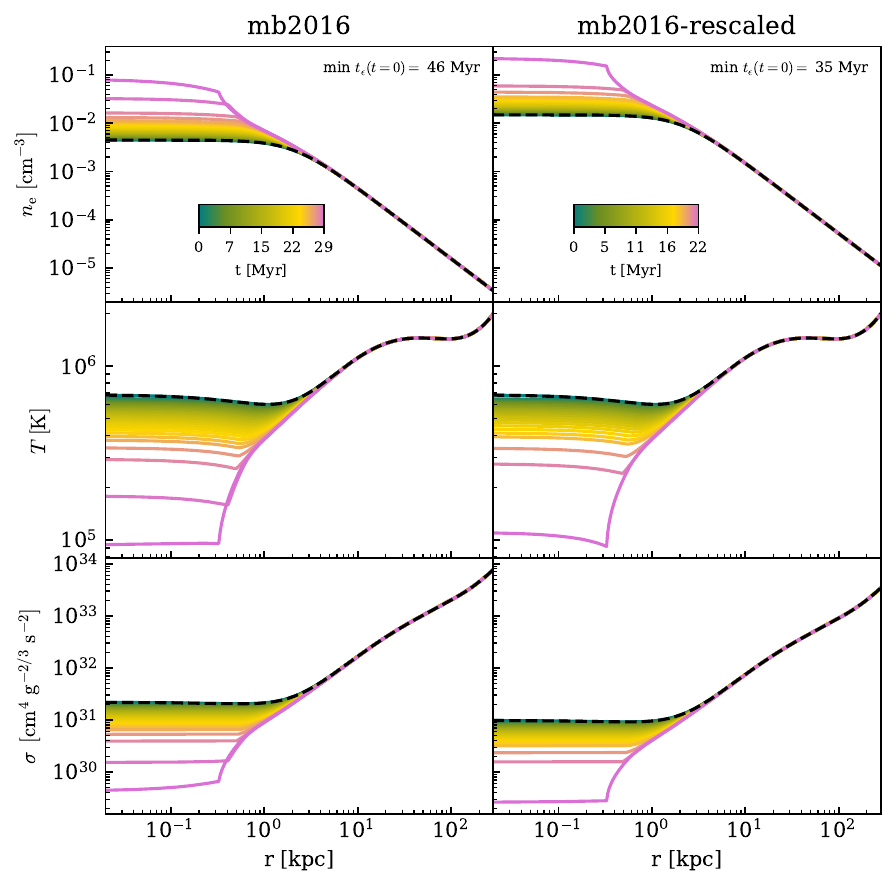}
	\end{center}
	\caption{
		Evolution of the electron density (top), temperature (middle) and
		entropy-index (bottom) for two initial conditions \mbnormal~(left) and
		\mbrescaled~(right). Each curve represents the \ac{CGM} at a specific
		time in the evolution. All curves are separated by
		\SI{1}{Myr} and time is indicated by the colour of the curve. The
		black dashed line highlights the values of the \ac{CGM} at $t=0$. The
		shortest characteristic time-scales
		(Equation~\ref{eq:characteristic-timescale}), minimized over radius
		at $t=0$, are reported for both initial conditions on the top panel of each
		column.  These
		results display the evolution in a purely cooling scenario (no
		mechanical heating and no photoionization).
	}\label{fig:results-compare-ic}
\end{figure*}

As an initial application of our evolutionary model to the Milky Way, we let
the hot CGM evolve through radiative cooling only. This will be analogous to
the original application of the model by \citet{kaiser2003SimpleModelsCooling},
who used it to estimate the time until the next \ac{AGN} heating event would
occur in the Hydra cluster. For this scenario we compute the cooling function using Cloudy, without
any radiative background that would photoionize the gas nor any mechanical
heating. The resulting cooling function is close to \ac{CIE}. We note that our
cooling rates around $T = 10^{5-6}\,\si{K}$ are 2 to 6
times lower compared to the classical \ac{CIE} cooling function from~\citet{sutherland1993CoolingFunctionsLowDensity}. Their cooling rates at low metallicity
($[\rm Fe/H] <-1.0$) are computed using primordial abundances, which are oxygen enhanced compared
to the relative abundances based on the solar values from~\citet{asplund2009ChemicalCompositionSun} used in this work.

Figure~\ref{fig:results-compare-ic} shows the evolution of the electron density
(top row), temperature (middle row) and entropy-index (bottom row) as a function of radius
for both the \mbnormal~(left column) and \mbrescaled~(right column) initial conditions. The
evolution of the CGM is indicated by the different curves, with colour
indicating the time of the evolution, whereas the black dashed line corresponds
to $t=0$. As expected in a purely cooling scenario, we see that entropy is
radiated
away and as a consequence decreases. This is paired with a decrease in temperature and an increase in the
electron density to maintain hydrostatic equilibrium. For both initial conditions,
we see that the evolution is dominated by the cooling in the inner region ($r <
	\SI{3}{kpc}$) of the \ac{CGM}, whereas its properties barely seem to evolve beyond \SI{10}{kpc}.

In this inner region, because of our initial condition, the gas has the highest
density, and the temperature is in the range where the cooling curve peaks;
consequently the cooling rate will be the highest. Due to cooling, the gas in
the inner region increases in density and decreases in temperature promoting an even
higher cooling rate. Since in a purely cooling scenario, there is no mechanism
to stop this cooling, the density will keep increasing and therefore the cooling
reaches a runaway state. In such a runaway state, the time-step of the evolution
decreases rapidly. At this time, the hydrostatic assumption is no longer
valid and we stop the evolution. For the \ac{CGM} starting with initial
condition~\mbnormal, this occurs after \SI{29}{Myr} and for the \mbrescaled~CGM
after \SI{22}{Myr}. The value of the shortest cooling time, minimized over radius of
both initial conditions ($t_\epsilon$), shown on the top panels of each column, is of the
same order ($t_{\rm end}\approx0.63t_{\epsilon}$) at these evolutionary times.
Both evolution times of the system are very short compared to the expected age of the
Milky Way hot \ac{CGM}, which is around
\SI{10}{Gyr}~\citep{vandevoort2011DropCosmicStar,gronnow2024DensityMilkyWays}. In what follows, we
investigate whether feedback from the Milky Way can provide enough heating to
avoid catastrophic cooling of the gas in the central region.

The main difference between the two initial conditions can be seen in the total
evolution times, where the \mbrescaled~\ac{CGM} evolves slightly faster than
the \mbnormal~\ac{CGM}. This can be explained by the fact that the density of
the rescaled initial condition is a factor of three higher than~\mbnormal, and
therefore cooling proceeds faster. The metallicity of the rescaled initial condition
is also a factor of three lower relative to the lower mass configuration. Although radiative cooling at these temperatures is
dominated by metal lines and this lowered metallicity decreases the cooling rate, the effect is more than compensated for by the increase due to higher density. The central density at the end of the
evolution of the \mbrescaled~\ac{CGM} is slightly higher than the
\mbnormal~\ac{CGM}. This is explained by the lower metallicity: at the same
density, the characteristic timescale $t_\epsilon$ is longer for a lower
metallicity, so the lower limit in the time-step is achieved at a higher
density.

Overall, the qualitative results from the evolution are identical for both
initial conditions. For the rest of the discussion, we therefore mostly focus
on the \mbnormal~\ac{CGM}.

\subsubsection{Mass inflow rate}\label{sec:results-inflow-rate}%

We showed that, in the pure cooling scenario, the \ac{CGM} beyond \SI{10}{kpc} hardly changes in
density, temperature or entropy. To investigate if any meaningful change is
occurring in this regime, we compute the rate of change in the mass at each
radius $\dot M(r)$. Since the change in mass at each radius can be over 5
orders of magnitudes smaller than the total mass of the system, small
uncertainties in the integration can dominate the value of $\dot M(r)$. To
suppress these fluctuations,  we therefore compute the change mass in interval
$t$ to $t + \Delta t$, where $\Delta t$ is large enough to contain at least
three evolutionary time-steps. We then take all the masses at $r$ in this
interval: $M_t(r),M_{t+ \delta t}(r), M_{t+ 2\delta t}(r) + \ldots M_{t+ \Delta
			t}(r)$. Assuming the increase in mass over time is roughly linear in this
interval, we then fit a line through these masses to obtain $\dot M_t(r)$. In
practice, we typically set $\Delta t\geq \SI{1}{Myr}$ such that 10 to 100
time-steps are included in a single fit. A positive value for $\dot M(r)$
indicates that mass is flowing towards the centre of the system, and we define
this quantity as the inflow rate at a specific radius $r$.

\begin{figure*}
	\begin{center}
		\includegraphics{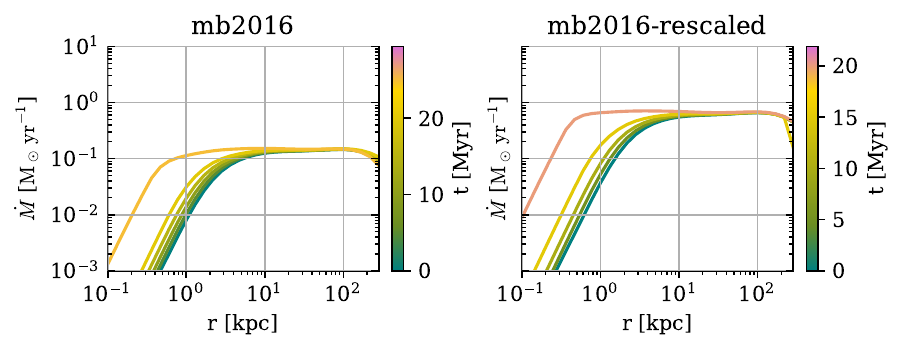}
	\end{center}
	\caption{
		Evolution of the change in mass, $\dot M(r)$, at each radius (see
		Section~\ref{sec:results-inflow-rate}) for two initial conditions
		\mbnormal~(left) and \mbrescaled~(right), without any heating. The
		colour of the curve indicates the time of the evolution. A positive
		value for $\dot M$ implies mass is flowing towards the centre of the
		system. $\dot M(r)$, in this figure, has been fitted using the
		interval $\Delta t = \SI{5}{Myr}$.
	}\label{fig:results-compare-mass-accretion-cooling}
\end{figure*}

Figure~\ref{fig:results-compare-mass-accretion-cooling} shows the evolution of
the inflow rate, with time indicated by curve colour, for both the
\mbnormal~(left) and \mbrescaled~(right) initial conditions. In both cases we
see that at very low radii, the inflow rate is close to zero and increases
rapidly with radius up to roughly a radius of few \si{kpc}. In this region, the
density is highest, but the total volume is small and therefore the change in
mass is also quite small. Beyond this radius, the inflow rate is relatively
constant for the majority of the \ac{CGM}, only decreasing slightly at the very
outer radius, where the mass is fixed due to our boundary condition. Over time,
the inflow rate changes mostly in the centre of the system, where we also saw
the
largest increase in density. Beyond a few \si{kpc}, the inflow rate is approximately
constant, reaching values of $\dot M \sim
	0.1-\SI{0.2}{\Msolpyear}$ and $\dot M \sim 0.6-\SI{0.7}{\Msolpyear}$ for the
\mbnormal~and \mbrescaled~\ac{CGM}, respectively. We emphasize that these values
have been obtained in absence of any heating.
The difference between the two initial conditions can be explained by the
different cooling rates of the gas as discussed in the previous section.

\citet{miller2013StructureMilkyWays} estimate the current mass accretion rate
of the Milky Way's hot \ac{CGM} by integrating spherical mass shells, divided
by the cooling time at the respective radii. At a radius of \SI{40}{kpc} (where
the cooling time is approximately equal to the Hubble time) they obtain a mass
accretion rate between 0.1 and \SI{0.2}{\Msolpyear}, for a density and
metallicity corresponding to the \mbnormal~\ac{CGM}. We see that at this
radius, our inflow rates for the \mbnormal~profile are compatible with the
values that they obtain.

The current \ac{SFR} of the Milky Way has been estimated to be between
$\SI{1}{}-\SI{3}{\Msolpyear}$~\citep{bland-hawthorn2016GalaxyContextStructural}.
The inflow rates we find in both configurations of the pure cooling scenario, assuming they reach the Milky Way disc to form
stars, are below \SI{1}{\Msolpyear}, and consequently they would not be
sufficient to fully sustain the current \ac{SFR}. The inflow rate of the
\mbrescaled~\ac{CGM} is close to the lower limit, however, and in practice the
amount of gas accretion needed to sustain star formation is slightly lower than
the \ac{SFR} \citep{fraternali2012EstimatingGasAccretion}. It is therefore
still possible that spontaneous cooling of the gas from the hot \ac{CGM} is a major
contributor to the total gas accretion. If the \mbnormal~initial condition is a
more realistic \ac{CGM}, we would need additional sources of gas accretion to
fully sustain the formation of stars in the Milky Way. To fully determine how much gas accretes from the hot \ac{CGM}, we would have to run the model for much longer. Moreover, these mass accretion rates have been obtained in the absence of any heating mechanisms, which can cause the accretion rates to decrease or for the cooling to cease altogether (see also Section~\ref{sec:comparison-with-fermi-bubbles}).

\subsection{Photoionization}\label{sec:fiducial-model}%

We now examine the evolution of a hot \ac{CGM} that in addition to radiative cooling is also influenced by
heating terms that photoionize the gas. We assume that the photoionizing
flux is due the the presence of an  extragalactic \ac{UVB}~\citep[$z=0$ values taken from][]{haardt2012RadiativeTransferClumpy}, star-forming regions in the Milky Way
(Section~\ref{sec:photoionization-from-star-forming-regions}) and
activity from \sgrAs~(Section~\ref{sec:photoionization-from-agn}).

\begin{figure*}
	\begin{center}
		\includegraphics{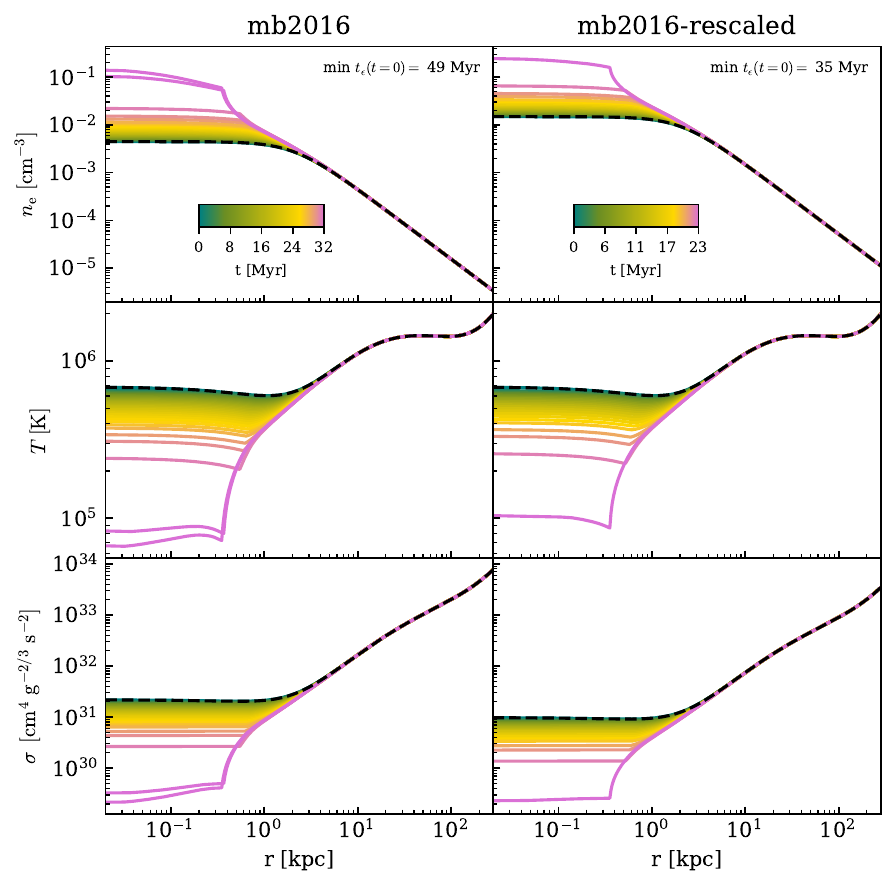}
	\end{center}
	\caption{
		The same as Figure~\ref{fig:results-compare-ic}, but with photoionization
		from the extragalactic \ac{UVB}, star-forming regions in the Milky Way and activity from \sgrAs.
	}\label{fig:results-compare-pi-ic}
\end{figure*}

Figure~\ref{fig:results-compare-pi-ic} shows how the two initial conditions
(left \mbnormal~and right \mbrescaled) evolve due to radiative cooling and
photoionization. In both cases, we see again that the evolution is dominated by
rapid cooling in the inner few \si{kpc}, where a high density (and low
temperature) is reached within \SI{32}{Myr} for the \mbnormal~and \SI{23}{Myr}
for the \mbrescaled~\ac{CGM}. The result of adding the photoionizing fields is
therefore a small time delay ($3$ and \SI{1}{Myr} for \mbnormal~and
\mbrescaled, respectively, compared to the scenario with only cooling) until a
runaway cooling state is reached in the hot \ac{CGM}. Again, the evolutionary
times are of the same order of the characteristic time-scale ($t_{\rm end}\approx
	0.65t_\epsilon$). It is therefore clear from these results that photoionization
does not provide enough heating to stop the gas from cooling in the centre.

For both initial conditions, we see that the constant inflow rates beyond
\SI{10}{kpc}, seen in the pure cooling scenario
(Section~\ref{sec:results-inflow-rate}), are unchanged by the addition of
heating terms that photoionize the gas (shown in Appendix~\ref{app:figures},
Figure~\ref{fig:results-compare-mass-accretion-pi}).
So, also beyond \SI{10}{\kpc}, we see that the
photoionization does not have a big impact on the cooling of the hot \ac{CGM}.

\subsection{Mechanical Heating}\label{sec:heating}%

From our previous results, it is apparent that a hot CGM that evolves only
through radiative cooling cools rapidly in the centre. Moreover, we see that
heating of the gas through photoionization due to the \ac{UVB} and local sources in the Milky
Way is also not enough to stop the central regions of the CGM from cooling rapidly. We now
turn on a central source that injects thermal energy into the
system. This injection is most representative of mechanical feedback from a
generic source in the centre of the system, implying that we do not explicitly link it to \ac{AGN} activity or  supernovae. This injection is controlled via two
parameters, $\dot E_{\rm tot}$, the total energy that is injected per unit time
(or the power), and $\lengthscale$, the length-scale of the Gaussian kernel over
which the energy is deposited (see
Equation~\ref{eq:mechanical-energy-injection}).

\begin{figure*}
	\begin{center}
		\includegraphics{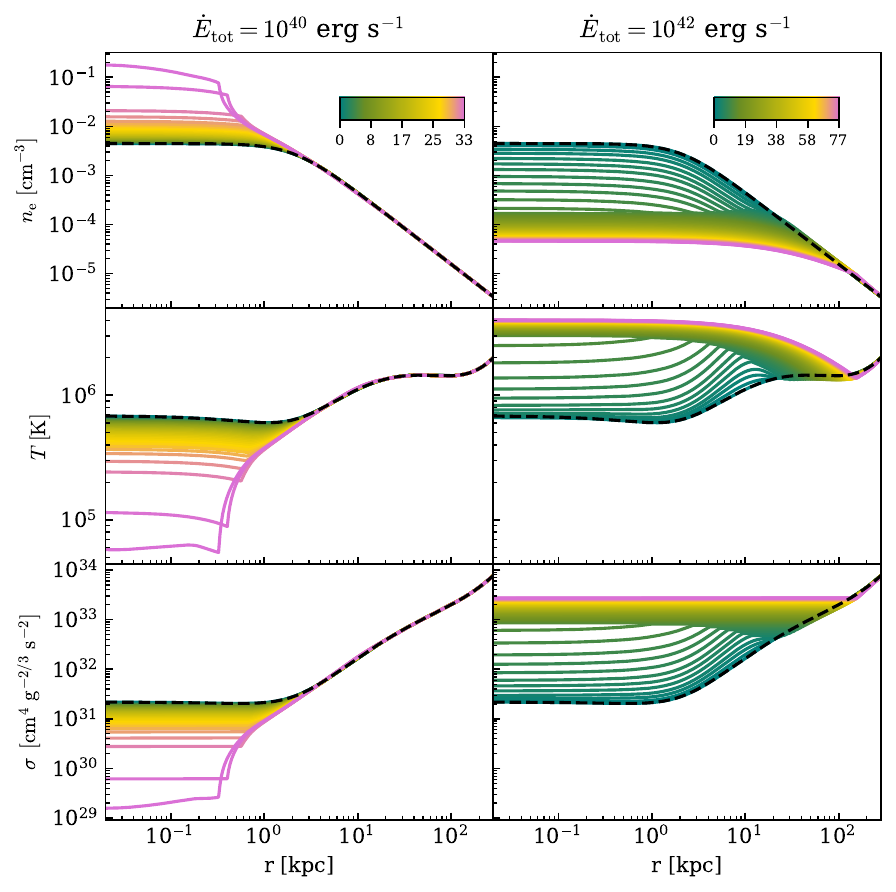}
	\end{center}
	\caption{
		The same as Figure~\ref{fig:results-compare-ic}, but with mechanical
		heating injected in a kernel of length-scale $\lengthscale = \SI{8}{kpc}$ and photoionization
		included. The two columns illustrate the evolution of the
		\mbnormal~\ac{CGM} with different amounts of mechanical heating. The
		left column shows the case where cooling is still dominant ($\dot
			E_{\rm tot} = \SI{e40}{\erg\per\s}$) and right a case where heating is
		dominant ($\dot E_{\rm tot} = \SI{e42}{\erg\per\s}$). Due to the
		different time-scales of the evolution, the curves in the left and right
		column are separated by \SI{1}{} and \SI{10}{Myr}, respectively.
	}\label{fig:results-compare-heating}
\end{figure*}

Figure~\ref{fig:results-compare-heating} shows the evolution of the
\mbnormal~\ac{CGM} which includes photoionization, but now also
mechanical heating from a source with length-scale $\lengthscale =
	\SI{8}{kpc}$. The two columns display the results for two extremes
of the total energy that is injected, $\dot E_{\rm tot}$. On the
left, we see a relatively low power ($\dot E_{\rm tot} =
	\SI{e40}{\erg\per\s}$) source, and as a consequence, the evolution proceeds similarly
to what we see in our previous results without mechanical heating
(Figure~\ref{fig:results-compare-pi-ic}, left column). For such a low power
we can therefore conclude that the \ac{CGM} evolution is dominated by cooling.
In the right column of Figure~\ref{fig:results-compare-heating}, we show the
evolution of a system where a high power ($\dot E_{\rm tot} = \SI{e42}{\erg\per\s}$) mechanical heating source is considered. In this case, the evolution is completely different from
what we discussed in a cooling dominated \ac{CGM}. In the first tens of \si{Myr}, the
central heating causes the \ac{CGM} within \SI{10}{kpc} to heat rapidly,
decreasing (increasing) the density (temperature) by an order of magnitude. At
around \SI{100}{Myr}, the evolution proceeds more slowly, but the source  progressively heats the \ac{CGM} even beyond \SI{100}{kpc} at the end of the
simulated time span. Similar to the case where cooling was dominant, the
temperature in the \ac{CGM} will have a runaway increase. The system could
continue to evolve for a longer time, but this would not change the runaway
state of the system. We therefore stop the evolution based on the number of
iterations and note that the evolutionary time of the system in this case does
not have a physical significance.

The entropy index (bottom row) also increases rapidly at the start of the
evolution, creating a large region in the system that is completely flat. This
flattening of the entropy is a result of how we implement convection (see
Section~\ref{sec:negative-entropy-gradients}): the injection of thermal energy
causes a negative entropy gradient in the centre, which is then flattened by
our convection scheme and as a result energy is transported to larger radii.

For the \mbrescaled~\ac{CGM}, the qualitative trend of low power being cooling
dominated, and high power being heating dominated is similar. We discuss the
main differences between the evolution of the two initial conditions in the
following section (Section~\ref{sec:balancing-heating-and-cooling}), where we
investigate how much energy injection is needed to change from a cooling to a
heating dominated \ac{CGM}. The length-scale also has an impact on this, but
again, the qualitative results remain the same (see
Section~\ref{sec:length-scale}).

\subsection{Balancing heating and cooling}\label{sec:balancing-heating-and-cooling}%

\begin{figure*}
	\begin{center}
		\includegraphics{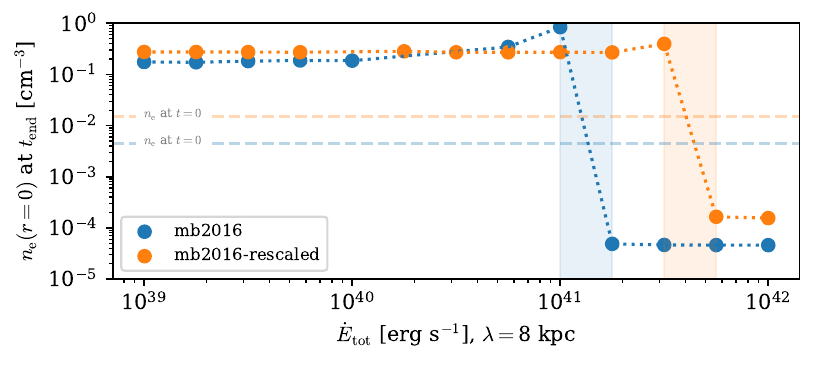}
	\end{center}
	\caption{
		Impact of mechanical heating coming from the centre of the system on the
		cooling of the gas in the hot \ac{CGM}. Each dot shows the electron density
		at $r=0$, at the end of the evolution with varying mechanical heating as
		indicated by the horizontal axis and length scale $\lengthscale = \SI{8}{kpc}$.
		The two initial conditions are indicated by colour where \mbnormal~and
		\mbrescaled~correspond to the blue and orange data points respectively. The
		shaded regions correspond to the region in which the \ac{CGM} transitions
		from a cooling- to a heating-dominated regime. As a reference, the electron
		density at $t=0$ is included as a dashed horizontal line.
	}\label{fig:final-density-S8}
\end{figure*}

The results in the previous section show that, depending on the amount of
mechanical energy that is injected, the hot CGM of the Milky Way can be
dominated by cooling (heating) for a low (high) $\dot E_{\rm tot}$. Both
scenarios are not realistic if one expects the \ac{CGM} to be relatively stable
for time-scales longer than the evolutionary times of a few tens of \si{Myr} presented
here. Between the extreme cases
presented in Figure~\ref{fig:results-compare-heating} we expect the evolution
in our model to transition between cooling and heating dominated regimes.

Figure~\ref{fig:final-density-S8} shows the result of several evolutionary runs
with varying total injected power shown on the horizontal axis. Each data point
corresponds to the electron number density $n_{\rm e}$ at $r=0$ at the end of
the evolution of an individual run, for both initial conditions indicated
by blue (\mbnormal) and orange (\mbrescaled). As a reference, the corresponding
$r=0$ electron densities at $t=0$ are included as horizontal dashed lines. This figure
shows the trend discussed in Section \ref{sec:heating}, namely that at
low $\dot E_{\rm tot}$ the electron density in the centre increases and the evolution is
therefore cooling dominated, whereas at high power the density decreases as a result of dominant heating.

Between these two extreme regimes, there is a sharp jump from high to low
electron densities as power increases. This region (shown as a shaded region)
indicates that, in this range of powers, the evolution transitions from cooling
to heating. We define the power at which this transition occurs as
\textit{threshold power}: around that value the cooling and heating in the
\ac{CGM} are balanced. For a mechanical heating source with $\lengthscale
	= \SI{8}{kpc}$, we find that the threshold power is between
$10^{41-41.25}$ and $10^{41.5-41.75}\si{\erg\per\s}$ for the \mbnormal~and
\mbrescaled~\ac{CGM}, respectively. The threshold power in the
\mbrescaled~\ac{CGM} is higher compared to the \mbnormal~\ac{CGM}, which is
explained by the higher cooling rate in the former.
We note that the stopping criterion varies between the cooling and heating
dominated regimes and therefore cannot be compared physically. However, in both
cases the central density would only diverge further from the intial value.
Therefore, changing the stopping criterion would not affect the location of the
jump and the corresponding threshold power.

A current limitation of the model is that our mechanical heating source injects
energy continuously. Because of
this, the temperature of the \ac{CGM} will increase in runaway fashion whenever the heating
dominates over the cooling. This is what causes the transition between the two
regimes to be so sharp: just below the threshold power, cooling will be
dominant and just above heating will be dominant and, in both cases, the system will be accordingly in a runaway cooling or heating state. In Section~\ref{sec:comparison-with-fermi-bubbles}, we compare
the threshold powers that we find to a more physical scenario, where the
central source operates in shorter episodes (or duty cycles).  In
Section~\ref{sec:comparison-with-supernovae}, we estimate whether sustained
supernovae in the \ac{CMZ} can supply enough heating to balance the hot \ac{CGM} cooling.

\subsubsection{Length scale}\label{sec:length-scale}%

Besides varying the total power that can be injected, we also change the
length-scale over which the power is distributed. Above, we have taken $\lengthscale
	= \SI{8}{kpc}$ to roughly match the size of the Fermi
bubbles~\citep{su2010GIANTGAMMARAYBUBBLES,predehl2020DetectionLargescaleXray}.
To test what happens for different length-scales, we evolved the same
\ac{CGM} profiles for the same range of $\dot E_{\rm tot}$ for $\lengthscale = \SI{5}{}$
and $\SI{2}{kpc}$. The results are shown in Appendix~\ref{app:figures},
Figures~\ref{fig:final-density-S5} and~\ref{fig:final-density-S2}.

The overall trends are the same, where there is a clear transition
between the cooling and heating regimes at low and high powers. For smaller
length-scales however, the threshold powers also decrease. For $\lengthscale =
	\SI{5}{kpc}$ they are $10^{40-40.5}\si{\erg\per\s}$ (\mbnormal) and
$10^{41-41.25}\si{\erg\per\s}$ (\mbrescaled). For $\lengthscale = \SI{2}{kpc}$,
the values are even lower: $10^{39.25-39.5}\si{\erg\per\s}$ (\mbnormal) and
$10^{40-40.25}\si{\erg\per\s}$ (\mbrescaled). A lower length-scale will result
in a distribution of the injected energy that has a higher concentration, and
therefore will inject more energy at $r=0$, for the same $\dot E_{\rm tot}$.
The threshold powers are based on the cooling and heating balance in the centre
of the system, and consequently the lower length-scales will give a lower
threshold power. We note that even though the \ac{CGM} might be dominated by
heating in the centre, the outer regions can still cool. In the discussion
(Section~\ref{sec:idealized-heating-case}), we elaborate on this by
comparing the threshold powers given by the different length-scales to
lower limits on the injected energy.

\subsection{Comparison with Fermi bubbles}\label{sec:comparison-with-fermi-bubbles}%

Our previous results show that the hot \ac{CGM} cools very rapidly close to the centre
of the system, and that this cooling cannot be stopped by photoionizing sources
in the galaxy. We then find the amount of energy that needs to be injected (that we define as the threshold power) in the centre of the system, in order to
stop the rapid cooling of the \ac{CGM} gas. In this section, we investigate
whether the source in the Milky Way's \ac{GC} that produced the Fermi
Bubbles~\citep{su2010GIANTGAMMARAYBUBBLES,predehl2020DetectionLargescaleXray}
could produce enough power to stop the rapid cooling of the hot \ac{CGM}.

\subsubsection{Potential sources that created Fermi bubbles }\label{sec:estimates-for-fermi-bubble-sources}%

The exact source that created the $\gamma$-ray emitting Fermi
bubbles~\citep{su2010GIANTGAMMARAYBUBBLES} and the larger X-ray emitting
eROSITA bubbles~\citep{predehl2020DetectionLargescaleXray} is still
unknown~\citep{sarkar2024FermiEROSITABubbles}. The two most likely candidates
are a short active phase of \sgrAs~\citep{yang2018UnveilingOriginFermi}, or
more sustained \ac{NSF} in the \ac{GC}~\citep{crocker2015UnifiedModelFermi}.
For our comparison, we will take several candidate sources directly based on
observations of the Fermi and eROSITA bubbles from
\citet{miller2016INTERACTIONFERMIBUBBLES}
and~\citet{predehl2020DetectionLargescaleXray}, and one simulation-based
source that produces the bubbles through the formation of a \ac{AGN}
jet~\citep{yang2022FermiEROSITABubbles}.

The candidate source from \citet{miller2016INTERACTIONFERMIBUBBLES} is based on
X-ray absorption in background \ac{AGN} sightlines close to the \ac{GC} which constrain
properties of both the hot \ac{CGM} and gas in the Fermi bubbles. They estimate
what could have inflated the bubbles by assuming a continuous Galactic outflow
following the energy conserving phase of
the~\citet{weaver1977InterstellarBubblesII} solutions. Using the observed
density and expansion velocity, they estimate the age of the bubbles to be
$t_{\rm age} = 3-\SI{5}{Myr}$ and the energy injection for both bubbles to be
$(1-7)\times\SI{e42}{\erg\per\s}$.

Using the all-sky X-ray emission maps from eROSITA,
\citet{predehl2020DetectionLargescaleXray} show the existence of an X-ray
emitting bubble surrounding the Fermi bubbles and derive the thermal properties
of the gas in this bubble. Assuming the sharp boundary of the bubble is due to
an adiabatic shock, they find an expansion velocity of roughly
\SI{340}{\km\per\s}. From this, they estimate that the expansion time of such a
bubble is roughly \SI{20}{Myr}, and has been inflated by a source with power
$(1-3)\times\SI{e41}{\erg\per\s}$. Alternatively, they speculate the creation
of the bubbles would also be compatible with a short ($1-\SI{2}{Myr}$) episode
of Seyfert-like activity with power \SI{e43}{\erg\per\s}.

\citet{yang2022FermiEROSITABubbles} take a different approach, and simulate a
short outburst with a jet produced by \sgrAs. They set the total power of each
jet to \SI{3.16e44}{\erg\per\s}, turn it on for only \SI{0.1}{Myr}, and
then let the system evolve for a total of \SI{2.6}{Myr}. As a result, both the
morphology and the sizes of the Fermi and eROSITA bubbles are reproduced.

\subsubsection{Estimating the average power of the candidate sources}\label{sec:average-power-of-the-estimates}%

The various scenarios for producing the Fermi bubbles described above show a wide
range in power ($\dot E_{\rm on}$) and time for which they are active
($t_{\rm on}$). Since all of these sources need to reproduce the Fermi bubbles,
which have a fixed total energy, the power of the source and $t_{\rm on}$ are
typically inversely related to each other: a high-power source needs less time
to inject the same amount of energy compared to a low-power source.

It is known that the feedback process of \ac{AGN} occurs not as a single
event, but rather in duty cycles. Typically, the \ac{AGN} spends only a short time in
an `outburst' state \citep[less than a percent of the total
	cycle,][]{cimatti2019IntroductionGalaxyFormation}, and is inactive for the
remainder of the time.

The average power that is injected by a source operating in such duty cycles is
given by
\begin{equation}
	\left<{\dot E}\right> = \frac{\dot E_{\rm on} t_{\rm on} }{t_{\rm on} + t_{\rm off}}
	\label{eq:duty-cycle}
\end{equation}
where $t_{\rm off}$ is the time in which the source is not injecting energy (i.e.
$\dot E_{\rm off} = 0$). We can calculate this average power for each of the
source estimates by plugging in their values for $\dot E_{\rm on}$ and $t_{\rm
			on}$. The time for which the source is off is not known, and therefore we
estimate lower and upper limits for $t_{\rm off}$. For the lower limit, we
assume that the next outbursts starts now, and
for the upper limit we adopt the cooling time of the eROSITA bubbles
(approximately $\SI{200}{Myr}$\footnote{
	We have performed the following experiment using our own model, to verify
	that this number is indeed representative for $t_{\rm off}$. Starting from
	an evolution that is dominated by the central mechanical heating (for
	example the right panel in Figure~\ref{fig:results-compare-heating}), we
	identify the time-step when a flat entropy core extending out to
	\SI{8}{kpc} forms, stop the mechanical heating and let the \ac{CGM} cool.
	For the \mbnormal~and \mbrescaled~\ac{CGM}, the time until a core of such
	size cools is \SI{220}{} and \SI{280}{Myr} respectively, which are
	reasonably close to the value of \SI{200}{Myr} that we adopt for $t_{\rm off}$.
}) estimated by~\citet{predehl2020DetectionLargescaleXray}. If a literature estimate includes a time-scale for when the emission has
stopped, we add this value to $t_{\rm off}$.
For~\citet{yang2022FermiEROSITABubbles} this is given to be \SI{2.5}{Myr}, and for the
Seyfert-like outburst of \citet{predehl2020DetectionLargescaleXray} we adopt
$18-\SI{19}{Myr}$, corresponding to the time needed to inflate the bubbles (\SI{20}{Myr}) after the short ($1-\SI{2}{Myr}$) outburst.

Calculating the average power of the sources allows us to make a comparison
with the threshold powers that we find. This is because the injection of energy
in our model is continuous, and therefore best represents an average energy
injection, assuming non-equilibrium effects are negligible. The value of the
threshold energy can then be interpreted as the average energy that is needed
to keep the \ac{CGM} stable for several \si{Gyr}. In Section~\ref{sec:non-equilibrium-effects} we discuss how non-equilibrium effects impact this comparison.
\begin{figure}
	\begin{center}
		\includegraphics{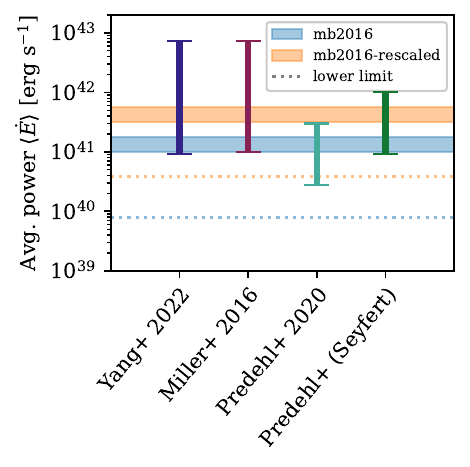}
	\end{center}
	\caption{
		Estimates of the average powers of the various scenarios that created the Fermi bubbles (from~\citealt{yang2022FermiEROSITABubbles},
		\citealt{miller2016INTERACTIONFERMIBUBBLES} and
		\citealt{predehl2020DetectionLargescaleXray}), compared with the threshold
		power that we find for both initial conditions \mbnormal~(blue) and
		\mbrescaled~(orange), for a central source with injection length-scale
		$\lengthscale=\SI{8}{kpc}$ (Figure~\ref{fig:final-density-S8}). The lower limits on the energy needed to balance
		the cooling (Section~\ref{sec:idealized-heating-case}) are indicated by dashed
		horizontal lines.
	}\label{fig:threshold-comparison-S8}
\end{figure}

The average power of the various candidate sources (x-axis), along with the threshold
powers for $\lengthscale = \SI{8}{kpc}$ (horizontal shaded area) are plotted in
Figure~\ref{fig:threshold-comparison-S8}, in which the two initial conditions are
indicated by blue (\mbnormal) and orange (\mbrescaled). The lower limits on the
energy needed to balance the cooling (see the discussion in
Section~\ref{sec:idealized-heating-case} for their derivation) are also included as horizontal
dotted lines. We first look at the average power of the various candidate sources. We note that the average power for the \citet{yang2022FermiEROSITABubbles}
and \citet{miller2016INTERACTIONFERMIBUBBLES} sources are very similar, whereas
the values of both sources from
\citet{predehl2020DetectionLargescaleXray} are considerably lower. This may be
attributed to the fact that~\citet{predehl2020DetectionLargescaleXray} only consider the X-ray component of the Fermi bubbles
and do not consider the $\gamma$-ray component.

When we compare our threshold values to the average energies for the Fermi
bubble sources, we find that they are strikingly similar. The values for both
the \mbnormal~and \mbrescaled~\ac{CGM} are within the lower- and upper bounds of
the average power estimated from the sources by~\citet{yang2022FermiEROSITABubbles},~\citet{miller2016INTERACTIONFERMIBUBBLES}
and the Seyfert-like burst by~\citet{predehl2020DetectionLargescaleXray}. Only
the upper limit of the \citet{predehl2020DetectionLargescaleXray} source is
below the threshold value of the \mbrescaled~\ac{CGM}. This indicates that the
source that created the Fermi bubbles can stop the rapid cooling of the
gas in the centre of the \ac{CGM}.

The exact values of our threshold power are sensitive to the adopted length-scale for injection, but they do not change our primary
results (see Appendix~\ref{app:figures},
Figure~\ref{fig:threshold-comparison-S5-2} for the cases
for $\lambda = 5 \, {\rm kpc}$ and $\lambda = 2 \, {\rm kpc}$). In general, a smaller length scale decreases our threshold value, whilst the
average Fermi bubble power remains the same. This implies that the source generating the Fermi
bubbles would be even more effective in stopping the rapid cooling of the \ac{CGM} gas. In
addition, all the average powers estimated from the Fermi bubble observations (except for
the~\citealt{predehl2020DetectionLargescaleXray} values) are well above the
lower limits  (dotted lines) derived in Section~\ref{sec:idealized-heating-case} for both initial conditions, further confirming
that the source that the created the Fermi bubbles is powerful enough to also balance the \ac{CGM} cooling.

It is noteworthy that the threshold values that we have derived for the hot
\ac{CGM} of the Milky Way, using a simple spherically symmetric model, are broadly in agreement with the average powers inferred observationally for the
generation of the Fermi bubbles. This agreement is consistent with a picture in
which the same source that inflated the Fermi bubbles can, averaged over its
duty cycle, offset radiative cooling in the inner hot \ac{CGM}, potentially
forming a self-regulating cycle that keeps the \ac{CGM} stable on \si{Gyr}
time-scales, longer than the model itself directly evolves.


\section{Discussion}\label{sec:Discussion} 
In this section, we will discuss the results that we obtained in the previous
section. In particular, we analyse how a different temperature profile in the
initial condition affects our result in
Section~\ref{sec:different-initial-condition}, discuss the impact of
non-equilibrium effects in Section~\ref{sec:non-equilibrium-effects} and compare
the threshold powers to idealized lower limits in
Section~\ref{sec:idealized-heating-case}. We then discuss whether supernova in
the \ac{CMZ} can halt the cooling of the \ac{CGM} in
Section~\ref{sec:comparison-with-supernovae}, and compare our model with cooling
flow models for galaxy-scale haloes in
Section~\ref{sec:comparison-cooling-flow-model}. Finally, we discuss various
model assumptions in Section~\ref{sec:model-assumptions}.

\subsection{Temperature and density profile in the initial condition}\label{sec:different-initial-condition}%

The results in the previous section have been obtained by assuming a
$\beta$-model density profile based on observations of the Milky Way's hot
\ac{CGM}~\citep{miller2016INTERACTIONFERMIBUBBLES} and a more massive and metal
poor variant to account for more recent observations in external
galaxies~\citep{zhang2024HotCircumgalacticMediuma}.
The first result we found is
that putting these density profiles in hydrostatic equilibrium with a \ac{NFW}
potential produces a temperature profile that is not isothermal (as assumed by
\citealt{miller2015ConstrainingMilkyWays}). Moreover, the resulting temperature
profile is distinctly different (black dashed lines in
Figure~\ref{fig:results-compare-ic}) from other models in the literature~\citep[for
	example see][Figure 2]{singh2024ComparisonModelsWarmhot}. The temperature in our
model hot \ac{CGM} increase with radius, up to the assumed value of
$10^{6.3}\,\si{\kelvin}$ at $r_{\rm vir}$, whereas in the literature the
temperature is most often seen to decline with radius.
In addition, it is likely that the hot CGM of the Milky Way has
been cooling for a long time, and it is possible that the density profile that
we observe today is a product of this cooling.

In our model, the potential is assumed to be static and therefore the
temperature in the centre is only sensitive to the core radius $r_{\rm c}$ of the
$\beta$-model (Equation~\ref{eq:beta-model}). As demonstrated by~\citet[][their
	Fig.~1]{guo2020ConstrainingMilkyWay}, a larger core radius will result in an
increased central temperature, but leaves the temperature profile beyond
\SI{100}{\kpc} identical. The temperature at these large radii is instead
completely determined by the value we assume at $r_{\rm vir}$ ($T(r_{\rm vir})=10^{6.3}\,\si{\kelvin}$). Unfortunately, observations have
difficulty constraining both the core radius and temperature at large radii in
the Milky Way halo (see Section~\ref{sec:initial-conditions-for-the-milky-way}).
A larger core radius, for the same normalization of the
$\beta$-model, produces a distribution with lower core density. Our results
show that the central density increases in a cooling \ac{CGM} (e.g.
Figure~\ref{fig:results-compare-ic}). Which indicates that a distribution with
lower central density could represent an earlier stage of the cooling Milky Way
\ac{CGM}.
In what follows, we show that changing the core radius of the initial density
distribution and the assumption of the temperature at the virial radius does not
affect our results significantly.

\begin{figure}
	\begin{center}
		\includegraphics{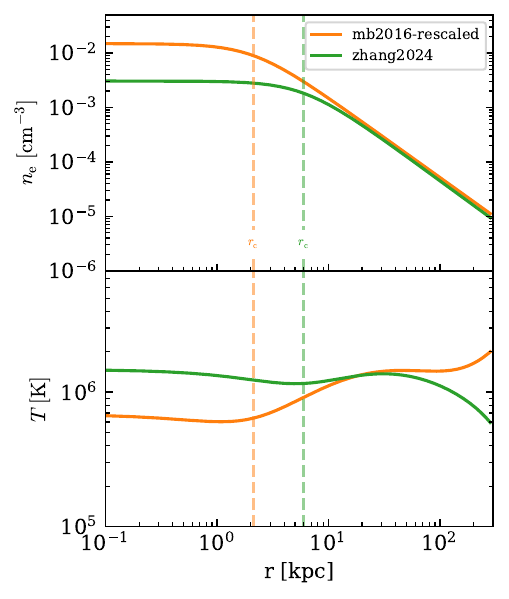}
	\end{center}
	\caption{Comparison between the density (upper panel) and temperature
		(lower panel) as a function of radius for the \mbrescaled~initial
		condition, and the initial condition described in
		Section~\ref{sec:different-initial-condition} (labeled as
		\texttt{zhang2024}). The dashed vertical lines indicate the value
		of $r_{\rm c}$ for the corresponding profile.
	}\label{fig:zhang2024-temperature-profiles}
\end{figure}

Following the same steps described in
Section~\ref{sec:initial-conditions-for-the-milky-way}, we create an initial
condition with a larger core radius of $r_{\rm c} =
	\SI{6}{\kpc}$\footnote{This is motivated by the value obtained
	by~\citet{zhang2024HotCircumgalacticMediuma} for their Milky Way-mass sample,
	although note that according to their uncertainties this value ranges from
	$\qtyrange{1}{15}{\kpc}$.} and a temperature at the virial radius equal to the
virial temperature of the halo $T(r_{\rm vir}) = T_{\rm vir,MW} \approx
	\SI{6e5}{\kelvin}$. The mass and metallicity are assumed to be identical to the
\mbrescaled~\ac{CGM}. Figure~\ref{fig:zhang2024-temperature-profiles} shows the
density and temperature profiles that result from these new parameters in
green, compared to the profiles from \mbrescaled~in orange. As expected, the
temperature in the centre is higher than the temperature of \mbrescaled,
whereas the temperature at the virial radius is lower. A small dip around the
core radius can still be seen, which would likely disappear for larger values
of $r_{\rm c}$

The evolution of this new initial condition (we include photoionization,
similarly to Section~\ref{sec:fiducial-model}), is qualitatively similar to the
evolution of the \mbrescaled~\ac{CGM} where cooling occurs mostly in the centre
of the system. With a higher temperature (lower density) in the centre, the
runaway cooling scenario does take significantly longer to develop
(\SI{275}{\Myr}), but this is still a relatively short time-scale compared to
the one for which we expect the \ac{CGM} to exist. Therefore, we still need some
additional heating source to balance this runaway cooling. Using the same method
outlined in Section~\ref{sec:balancing-heating-and-cooling}, we estimate that
the amount of energy needed for this \ac{CGM} to be in balance (the threshold
power) lies between $10^{40-40.5}\si{\erg\per\s}$. This value is significantly
below the one we obtain for the \mbrescaled~\ac{CGM}, but would not change the
conclusion that the source generating the Fermi bubbles can halt the cooling of
the \ac{CGM}, as shown in Section~\ref{sec:comparison-with-fermi-bubbles}.

The mass inflow rate (shown in Appendix~\ref{app:figures},
Figure~\ref{fig:discussion-mass-accretion-pi-zhang2024}) of this initial
condition shows some differences compared to our original models (shown in
Figure~\ref{fig:results-compare-mass-accretion-cooling}). In the inner
\SI{10}{kpc} the inflow rates, although developing on time-scales of
\SI{100}{Myr}, are qualitatively similar to the inflow rates of the
\mbrescaled~initial condition . At \SI{10}{kpc}, however, the accretion rate is
lower ($\dot M\approx0.2-\SI{0.5}{\Msolpyear}$) compared to the original rates ($\dot
	M\approx0.6-\SI{0.7}{\Msolpyear}$). Moreover, the accretion rate increases with
radius up to a maximum of $\dot M\approx\SI{0.9}{\Msolpyear}$ at \SI{100}{\kpc},
whereas the original rates stay constant. The discrepancy between
the two models beyond \SI{10}{kpc} can be explained by the decreasing
temperature of the new initial condition in this region. We therefore conclude
that constraining the temperature of the hot \ac{CGM} is crucial to
understanding its large-scale ($r\gtrsim \SI{10}{\kpc}$) mass inflow rates.

\subsection{Impact of non-equilibrium effects}\label{sec:non-equilibrium-effects}%

In Section~\ref{sec:comparison-with-fermi-bubbles}, we estimate the average
power that is injected during many duty cycles by several candidate sources
that created the Fermi bubbles. There are two main non-equilibrium effects that
could play a role in the estimation of the average powers emitted by the
sources. The first is the non-equilibrium state that results from a recent
change in the photoionizing
field~\citep{oppenheimer2013AGNProximityZone,segers2017MetalsCircumgalacticMedium,oppenheimer2018FlickeringAGNCan},
and the second is the non-equilibrium distribution of ions caused by rapid
cooling at temperatures below $T <
	\SI{e6}{\kelvin}$~\citep{sutherland1993CoolingFunctionsLowDensity,gnat2007TimedependentIonizationRadiatively}.

In this work, we include a photoionizing field coming from local Galactic
sources and the uniform \ac{UVB}. The \ac{UVB} will remain constant over the
time-scales on which the possible duty cycle of the  Fermi bubble source could
operate. The \ac{SFR} in the disc is also assumed to be relatively constant over
time-scales longer than the those considered in this work and therefore the
photoionizing field will not vary significantly. Moreover, the \ac{SFR} in the
\ac{CMZ} has been relatively constant for the past \SI{100}{Myr}, only
increasing from \SI{0.07}{} to \SI{0.09}{\Msolpyear} between \SI{5}{} to
\SI{100}{Myr} ago~\citep{henshaw2023StarFormationCentral}.~\sgrAs~does show
evidence of higher activity in the past~\citep{sarkar2024FermiEROSITABubbles}.
However, in Figure~\ref{fig:local-flux}, we see that at \SI{1}{kpc} the current
flux of \sgrAs~is five orders of magnitudes below the flux coming from
the disc. Since the luminosity of an \ac{AGN} is proportional to its mass
accretion rate (Equation~\ref{eq:agn-luminosity}), the accretion rate of
\sgrAs~would have to increase by the same amount in order to be relevant. Such
an increase in the photoionizing flux could be possible in the scenario of a short energetic episode, as
is considered by~\citet{yang2022FermiEROSITABubbles}. In this case, the cooling of the inner \ac{CGM}
might be affected due to the elevated ion distributions, but the exact effect
would require detailed calculations~\citep[see, e.g., ][]{oppenheimer2013AGNProximityZone,gnat2017TimedependentCoolingPhotoionized,segers2017MetalsCircumgalacticMedium,oppenheimer2018FlickeringAGNCan}
that are beyond the scope of this work.

Most likely a more significant effect is the non-equilibrium cooling that arises
when a gas, initially in \ac{CIE} or photoionization equilibrium, cools below
temperatures of $T<\SI{e6}{\kelvin}$. At these temperatures, the recombination
time-scales of the ions responsible for radiative cooling (for example,
\textsc{C iii} at $T=\SI{e5}{\kelvin}$) of the gas are significantly longer than
the cooling time of the gas causing a significant suppression in the cooling
rates~\citep{sutherland1993CoolingFunctionsLowDensity,gnat2007TimedependentIonizationRadiatively}.
It is noted that even by taking the non-equilibrium effects into account, the
gas at these temperatures is still prone to thermal
instabilities~\citep{sutherland1993CoolingFunctionsLowDensity}, which in
principle is what happens in the runaway cooling scenario in the centre of our
system. A slightly decreased cooling rate would affect the time it takes for the
Fermi bubbles to cool, and therefore increase $t_{\rm off}$ slightly. This would
not change the fact that the various estimates for the Fermi bubble source are
powerful enough to stop the cooling. In addition, a decreased cooling rate in
the centre of the \ac{CGM} would make it easier for a source to balance the
cooling, strengthening our conclusions.

\subsection{Idealized heating}\label{sec:idealized-heating-case}%

In Section~\ref{sec:balancing-heating-and-cooling} we have
obtained the amount of mechanical heating that is needed to stop the
hot \ac{CGM} from rapidly cooling in the centre. This value depends on the
shape over which we inject the feedback energy
(Equation~\ref{eq:mechanical-energy-injection}), since a smaller length-scale
$\lengthscale$ results in a smaller value for the threshold power
(Section~\ref{sec:length-scale}). An alternative approach to finding the
threshold power is to consider an idealized heating scenario, where the amount
of injected energy at each radius is exactly equal to the energy that will be
radiated away by cooling at that radius. Clearly this is not a physical
scenario, but it has the advantage that we do not have to assume a distribution over which energy is injected to estimate the amount of heating required to balance radiative losses. It will therefore represent a lower limit to the mechanical
heating required to offset the cooling of the \ac{CGM}.

The total net energy that is radiated away up to radius $r$ is given by
\begin{equation}
	\dot E(\leq r) = 4\pi\int_0^r \mathbb{L}(Z, T, n_{\rm H}, \Phi_{\rm SFR}, \Phi_{\rm AGN})\,{r^\prime}^2\dd r^\prime,
	\label{eq:power-radiated}
\end{equation}
where $\mathbb{L}_{\rm rad}$ is the net energy that is radiated away per unit
volume, and depends on the local properties of the gas and the incident
photoionizing flux evaluated at radius $r^\prime$. For both
initial conditions we compute the cooling radius $r_{\rm cool}$, the radius at
which the cooling time (Equation~\ref{eq:characteristic-timescale}) is equal
to the Hubble time ($t_{\rm H} \sim \SI{13.6}{\Gyr}$). For the \mbnormal~and
\mbrescaled~\ac{CGM}, the cooling radius is $r_{\rm cool} = \SI{52}{}$ and
\SI{65}{\kpc}, respectively. The total net energy that needs to be injected in
order to offset the radiative cooling up to this radius is then $\dot E(\leq
	r_{\rm cool}) = \SI{7.9e39}{\erg\per\s}$ for the fiducial density profile and $\dot E(\leq
	r_{\rm cool}) =\SI{3.85e40}{\erg\per\s}$ for the rescaled version.

Since these values are expected to be lower limits on the energy injection
needed to offset the cooling, we compare these with the threshold powers that
we obtained by varying the magnitude and length-scale of the mechanical heating
in the centre of the system (Section~\ref{sec:balancing-heating-and-cooling}).
For both initial conditions, the lower limits derived above are well below the threshold
powers obtained using the $\lengthscale = \SI{5}{}$ and \SI{8}{\kpc} length-scales.
Especially the distribution with \SI{8}{\kpc} is an order of magnitude above
the lower limit, which might indicate that the width of the Gaussian
distribution is too large. For some values of $\dot E_{\rm tot}$, a wide
distribution will cause the \ac{CGM} to heat around \SI{1}{\kpc}, whilst still
cooling at $r=0$. Since we estimate the threshold values by the density at
$r=0$ (Figure~\ref{fig:final-density-S8}), these cases will still be categorized as
cooling dominated.

The threshold powers for our smallest length-scale $\lengthscale=\SI{2}{\kpc}$
are instead slightly below the lower limit given by $\dot E(\leq r_{\rm cool})$.
This indicates that this distribution has a sharp peak, and produces a very localized
heating in the centre whilst the outer region is still cooling. Effectively,
this is the opposite scenario of the configuration we just examined, and will therefore be
interpreted as heating dominated in our estimation. If instead of using the
cooling radius, we calculate the lower limit using the radius where the cooling
time equals \SI{1}{Gyr}, we still obtain values that are above the threshold
powers for this length-scale. This indicates that the threshold value obtained
with the \SI{2}{\kpc} length-scale could be in an approximate equilibrium for
some time shorter than or comparable to a \si{\Gyr}, but will still cool on longer time-scales.
Therefore, we can conclude that injecting energy following a distribution with
length-scale $\lengthscale = \SI{2}{kpc}$, is too narrow to give an accurate
threshold power that balances the cooling of the hot \ac{CGM} gas on
time-scales longer than a \si{Gyr}. Results obtained with $\lambda = 5$ or \SI{8}{kpc} are therefore preferable.

\subsection{Can supernovae in the \ac{CMZ} stop the cooling of the \ac{CGM}?}\label{sec:comparison-with-supernovae}%

Our results demonstrate that a source that creates the Fermi bubbles, can stop
the rapid cooling of the inner hot \ac{CGM} and potentially form a
self-regulating cycle with the \ac{CGM} that keeps the \ac{CGM} of the Milky Way
stable for several \si{Gyr}. This result is independent of the nature of the
source that produces the power, as the argument is based on the total energy
budget. Studies on the nature of this source have demonstrated that the current
\ac{SFR} in the Galactic centre is too low for supernovae to provide enough
power to produce the Fermi bubbles~\citep{sarkar2024FermiEROSITABubbles}.
Therefore, the most likely candidate is a previous active phase of
\sgrAs~\citep{miller2016INTERACTIONFERMIBUBBLES,yang2018UnveilingOriginFermi,zhang2020SimulatingFermiBubbles}.
Even if supernovae in the Galactic centre are not powerful enough to produce the
Fermi bubbles, they could still provide enough energy to halt the rapid cooling
of the \ac{CGM}.

In a star forming region that is continuously forming stars with a given \ac{SFR}, the amount of mechanical feedback produced by Type II supernovae can be estimated by
\begin{equation}
	\dot E_{\rm SN,mech} = \epsilon E_{\rm SN, mech}\,N_{\rm SN}\,{\rm SFR},
	\label{eq:sn-mechanical-feedback}
\end{equation}
where $E_{\rm SN,mech} = \SI{e51}{\erg}$ is the mechanical energy of a single supernova, $N_{\rm SN}$ is the number of type-II supernovae for a star
forming region per unit stellar mass formed, and $\epsilon\approx0.3$ is the thermalization
efficiency~\citep{sarkar2015MultiwavelengthFeaturesFermi,miller2016INTERACTIONFERMIBUBBLES,cimatti2019IntroductionGalaxyFormation}.
To compute $N_{\rm SN}$ we assume
a~\citet{chabrier2003GalacticStellarSubstellar} \ac{IMF} and integrate between
stellar masses of $8$ and $\SI{300}{\solMass}$, giving $N_{\rm SN} = \SI{0.016}{\solMass\tothe{-1}}$.

The amount of mechanical feedback produced in the Galactic centre of the Milky
Way, using ${\rm SFR}=\SI{0.07}{\Msolpyear}$ for the
\ac{CMZ}~\citep{henshaw2023StarFormationCentral}, will then be roughly $\dot
	E_{\rm CMZ} \approx \SI{e40}{\erg\per\s}$. The values for the threshold
power, using length-scale of $\lengthscale = \SI{8}{kpc}$, are an order of magnitude
above the value of $\dot E_{\rm CMZ}$ and therefore mechanical feedback from supernovae in
the \ac{CMZ} alone would not be sufficient to stop the rapid cooling of the hot \ac{CGM}. However,
when we consider the threshold values with a smaller length-scale,
$\lengthscale=\SI{5}{kpc}$ for \mbnormal~and $\SI{2}{kpc}$ for both initial
conditions (see Figure~\ref{fig:threshold-comparison-S5-2}), the
mechanical feedback from supernovae would be sufficient to offset the cooling in
the centre.

A factor that we have not considered in this estimate is the star formation in
the disc of the Milky Way. An estimate of the disc's contribution to the
mechanical feedback requires us to consider geometrical effects, and
account for a more uncertain thermalization efficiency. Hence, this contribution
would be too uncertain to estimate. However, the contribution can only result in an increased value of $\dot E$ and therefore it could bring the value closer to
our threshold powers.

In Section~\ref{sec:idealized-heating-case}, we compute the idealized lower
limits on the energy that needs to be injected in the centre of the system for
the \ac{CGM} to be stable, resulting in $\dot E(\leq r_{\rm cool}) =
	\SI{7.9e39}{}$ and \SI{3.85e40}{\erg\per\s} for the \mbnormal~and
\mbrescaled~\ac{CGM} respectively. In this idealized heating scenario, the
mechanical feedback coming from supernovae in the \ac{CMZ} would be sufficient
to halt the cooling for the \mbnormal~\ac{CGM}, but not in the rescaled case.
However, considering the idealized lower limits as well as the value of $\dot
	E_{\rm CMZ}$ contain many uncertainties and the values are not separated that
much, we cannot completely rule out continuous feedback from supernovae in the
\ac{CMZ} as a mechanism to stop the cooling of the \ac{CGM}.

\subsection{Similarity with cooling flow models}\label{sec:comparison-cooling-flow-model}%

In Section~\ref{sec:results-inflow-rate} we show that our evolutionary model of
the hot \ac{CGM} predicts a constant mass inflow rate $\dot M(r)$ at radii
larger than \SI{10}{\kpc}. For the two initial conditions, these rates are $\dot
	M \approx \SI{0.1}{\Msolpyear}$ (\mbnormal) and $\dot M \approx \SI{0.6}{\Msolpyear}$
(\mbrescaled). Such large-scale inflowing hot gas, known as a cooling flow, has been
extensively studied in the context of haloes hosting galaxy clusters~\citep{mathews1978RadiativeAccretionFlow,fabian1984CoolingFlowsClusters,bertschinger1989EvolutionCoolingFlows}.
These cooling flow models predict inflow rates in the centre of the haloes
that were not matched by the observed \ac{SFR} in these central galaxies. This mismatch between theory and
observations was solved by offsetting the cooling through \ac{AGN} feedback
coming from the central galaxy in the
cluster~\citep{binney1995EvolvingCoolingFlows,omma2004StructuralStabilityCooling,mcnamara2007HeatingHotAtmospheres}. Following the approach
of~\mbox{\citet{fabian1984CoolingFlowsClusters}}, cooling flow models of
galaxy-scale haloes have been developed
by~\mbox{\citet{stern2019CoolingFlowSolutions}}.
In their model, the cooling
flow is fed continuously at a rate of $\dot M_{\rm flow}$, by either an outwards
growing cooling wave at the cooling
radius~\citep{bertschinger1989EvolutionCoolingFlows}, or gas that is
continuously accreted and shock heated from the \ac{IGM} onto the halo. The accreted gas
will cool slowly and flow towards the galaxy at subsonic speeds. Close to the
Galactic disc,
angular momentum becomes important, and the flow can be influenced by Galactic
feedback, therefore the model no longer applies in this regime. Assuming the
halo is dominated by a Galactic potential ($\Phi = -\int(v_{\rm c}^2/r)\dd r$)
with a circular velocity $v_{\rm c}$ that follows a power law $v_{\rm c} \propto r^m$,
\citet{stern2019CoolingFlowSolutions} obtain self-similar solutions for the
density, temperature and radial velocity of the inflowing gas throughout the
halo of the form
\begin{equation}
	\rho\propto r^{m-3/2}, \quad T\propto r^{2m}, \quad v_r\propto r^{-m-1/2}.
	\label{eq:stern2019-power-laws}
\end{equation}

To compare the inflow rate of our model with the cooling flow model, we assume
the density of the cooling flow model follows a $\beta$-model at large
radii\footnote{for $r \gg r_{\rm c}$ Equation~\ref{eq:beta-model} will follow
	$\rho\propto r^{-3\beta}$} with $\beta=1/2$, which is equivalent to assuming
$m=0$. We also note that this value is not too far from $m=-0.1$ adopted
by~\citet{stern2019CoolingFlowSolutions}. The full self-similar solution for
the radial velocity then reads
\begin{equation}
	\begin{aligned}
		-v_r(r)  = 12.9 \, & \pqty{\frac{M_{\rm vir}}{\SI{e12}{\solMass}}}^{-0.36}
		\pqty{\frac{\dot M_{\rm flow}}{\SI{1}{\Msolpyear}}}^{1/2}                           \\
		                   & \pqty{\frac{\Lambda}{\SI{e-22}{\erg\cm\tothe{3}\per\s}}}^{1/2}
		\pqty{\frac{r}{\SI{100}{\kpc}}}^{-1/2}\si{\km\per\s},
	\end{aligned}
	\label{eq:stern2019-inflow-velocity}
\end{equation}
where $M_{\rm vir}$ is the virial mass of the halo and $\Lambda$ the cooling rate
of the gas. The inflow velocity from our model can be computed from the continuity
equation, using the mass inflow rate
\begin{equation}
	v_r(r) = -\dot M(r) / 4\pi\rho(r) r^2,
	\label{eq:inflow-rate}
\end{equation}
where we note that the mass inflow rate of our model $\dot M(r)$ and
the accretion rate of the cooling flow model $\dot M_{\rm flow}$ do not have to
be equal.

\begin{figure}
	\begin{center}
		\includegraphics{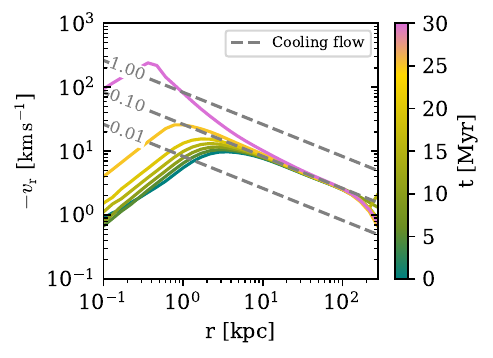}
	\end{center}
	\caption{
		Comparison between the inflow velocity of our evolutionary model and
		the solutions for a steady-state cooling flow
		from~\citet{stern2019CoolingFlowSolutions}. The coloured curves show the
		inflow velocity for the \mbnormal~\ac{CGM} at different times. The
		cooling flow solutions have been included as dashed grey lines for
		three different accretion rates $\dot M_{\rm flow} = \SI{0.01}{},
			\SI{0.1}{}$ and \SI{1.0}{\Msolpyear}, indicated in the Figure at
		the top left of the corresponding line.
	}\label{fig:cooling-flow-comparison}
\end{figure}

To compare the cooling flow solution with the evolution from the
\mbnormal~\ac{CGM}, we take $M_{\rm vir}=\SI{1.3e12}{\Msol}$ for the Milky
Way; moreover, since the cooling flow model assumes a constant cooling rate, we take
$\Lambda = \mathbb{L} \approx \SI{5e-23}{\erg\cm\tothe{3}\per\s}$ for our
cooling function at $T\sim \SI{e6}{\kelvin}$ with
metallicity $Z=\SI{0.3}{Z_\odot}$. Figure~\ref{fig:cooling-flow-comparison}
shows the inflow velocity of the \mbnormal~\ac{CGM} during the evolution of the
system, where the colour indicates the evolutionary time. The gray dashed lines
show the inflow velocity (Equation~\ref{eq:stern2019-inflow-velocity}) for mass
accretion rates of $\dot M_{\rm flow} = \SI{0.01}{}, \SI{0.1}{}$ and
$\SI{1.0}{\Msolpyear}$. The qualitative trend in inflow velocity of our model
is similar to the trend in the mass inflow rate in
Figure~\ref{fig:results-compare-mass-accretion-cooling}. At large radii the
inflow velocity decreases with radius due to the $1/\rho r^2$ factor, but stays
constant over time. When we compare the inflow velocity of the cooling flow
model with the inflow velocity from evolution of the \mbnormal~\ac{CGM} we see
that both models predict an identical slope. An even more striking similarity
between our two models is that the cooling flow model for $\dot M_{\rm flow}  =
	\SI{0.1}{\Msolpyear}$ is essentially identical to that of our evolutionary
model which, as we have seen before, has constant $\dot M$ between \SI{0.1}{}
and \SI{0.2}{\Msolpyear} at large radii. Below \SI{10}{kpc} the two models
differ significantly, but in this region the angular momentum of the cooling flow, as well as Galactic feedback becomes relevant and therefore the cooling
flow model no longer applies.

We emphasize that even though the fundamental equations governing the two
systems are similar, our model~\citep[based on ][]{kaiser2003SimpleModelsCooling}
and the cooling flow model from~\citet{stern2019CoolingFlowSolutions} make
completely different assumptions to obtain the properties of the hot \ac{CGM}.
Our evolutionary model assumes a closed system in approximate hydrostatic
equilibrium that slowly cools via radiative cooling, whereas the cooling flow
model assumes a continuous steady-state accretion process that determines the
properties of the \ac{CGM}. From this, we can conclude that at large radii our
evolutionary model is also representative for a halo that is being fed by an
outwards growing cooling wave, or by continuous accretion from the \ac{IGM}.
Conversely, we can conclude that the cooling
flow models for galaxy-scale haloes are close to hydrostatic
equilibrium.

\subsection{Model assumptions}\label{sec:model-assumptions}%

In this section, we briefly examine various assumptions underlying our model,
and show that they do not have an impact on our main results. We discuss our
treatment of convection, whether hydrostatic equilibrium is valid, and, finally,
discuss the robustness of our results if a different starting radius for the
integration is adopted . Most of the tests are performed on the
\mbnormal~\ac{CGM} that includes radiative cooling and photoionization
(Section~\ref{sec:fiducial-model}), and we refer to this model as our fiducial
model.

\subsubsection{Convection}\label{sec:convection}%

To avoid convectively unstable regions from forming due to the small negative
entropy gradient in the initial condition, we implemented a simple form of
convection (see Section~\ref{sec:negative-entropy-gradients}). In our model
runs without mechanical heating, the size of the convective region is at most
\SI{1}{kpc}, and the isothermal sound speed at $T\sim\SI{7e5}{\kelvin}$ is
$c_{\rm s} \approx \SI{100}{\km\per\s}$. We can then approximate the time-scale on which convection
acts as the size of the convective region $L$, divided by
the isothermal sound speed, giving $t_{\rm conv} \approx L / c_{\rm s} \approx
	\SI{10}{\Myr}$. This is two orders of magnitudes larger than our smallest
time-step. To test whether this affects our
results, we run our fiducial model with and without the convection scheme
enabled. In both cases, the evolution time of the system is identical and we
can conclude that it does not have an impact on the rapid cooling in the
centre. Moreover, our results concerning the mass inflow rates
(Section~\ref{sec:results-inflow-rate}) depend on the evolution of the \ac{CGM}
beyond \SI{10}{kpc}. Since the convective region does not reach
beyond \SI{1}{\kpc}, our results concerning the mass inflow rates are also
unaffected.

The mechanical heating case is slightly different. Indeed, in an overheating
scenario the convective region grows up to even \SI{100}{kpc} in extreme cases
(see Figure~\ref{fig:results-compare-heating}). Clearly, it would take a
significantly longer time-scale than the time-step of our model for convection
to act on such scales. However, considering that without the convection,
gas in the centre will heat to higher temperatures, this would imply that the
cooling rate in the centre would be less efficient, and in turn, would require
less mechanical feedback to offset. Therefore, the presence of convection would not change a heating-dominated scenario  into one that is dominated by cooling. On the other hand, if the
\ac{CGM} is dominated by cooling, the convective region will be smaller than
the length-scale of the kernel and therefore, including convection will not cause thermal
energy that is injected within the kernel to be transported away. So, a cooling-dominated case would also not be affected by the inclusion of convection. In particular, we
conclude that our treatment of convection does not influence the values we obtained
for the threshold power
(Section~\ref{sec:balancing-heating-and-cooling}).

\subsubsection{Hydrostatic equilibrium}\label{sec:hydrostatic-equilibrium}%

One of the underlying assumptions of our model is that at each time-step, the
hot \ac{CGM} is in hydrostatic equilibrium. To verify that the system has enough
time to reach such an equilibrium after a time-step, we compute the inflow
velocity $-v_{\rm r}$ (Equation~\ref{eq:inflow-rate}) at each radius for our
fiducial model and compare it to the isothermal sound speed at that point. The
inflow velocity at all radii is subsonic ($\abs{v_{\rm r}} / c_{\rm s} < 1$), so
the displacements caused by the radiative cooling are small enough such that the
\ac{CGM} has time to stay approximately in hydrostatic
equilibrium~\citep{forman1985HotCoronaeEarlyType}. Only at the end of the
evolution does the velocity become transonic ($\abs{v_{\rm r}} \sim c_{\rm s}$)
in the centre of the system, which would indicate hydrostatic equilibrium would
no longer hold there. Our model show that the \ac{CGM} reaches this point where
the gas is rapidly cooling, but it is unable to describe how this collapse
proceeds. To further study the properties of the \ac{CGM} beyond this time, more
sophisticated models of spherical collapse would be required.

\subsubsection{Inner radius}\label{sec:inner-radius}%

The numerical integration of Equations~\ref{eq:hydrostatic-equilibrium-entropy}
and~\ref{eq:mass-shells-entropy} requires an initial condition on the pressure
and mass. We find the pressure by an iterative procedure, and we assume the mass to
be $M=0$ at $r=0$. With these assumptions, we have to integrate the system, starting at
$r=0$, where density of the hot \ac{CGM} is not very well constrained. Instead,
we can assume that within some radius $r_0 > 0$, the mass, $M(r_0) = M_0$, is
constant in time, and use this as the boundary condition for the numerical
integration. The advantage is that we no longer have to consider what happens
below $r < r_0$, but because $M_0$ is constant, no mass can flow through the
surface at $r_0$ resulting in a pile up of mass at this radius. We evolve the
fiducial model with $r_0 = \SI{1}{\kpc}$, which results in a similar
evolutionary time, and in similar mass inflow rates at $r > r_0$ compared to the evolution
with $r_0=0$. This indicates that the evolution of the hot \ac{CGM} within
\SI{1}{\kpc} does not have an appreciable impact on the evolution at larger radii or on the general conclusions of our paper.


\section{Conclusions}\label{sec:Conclusions} 
In this work, we investigated the evolution of the hot \ac{CGM} around the
Milky Way by extending the semi-analytical model of \citet{kaiser2003SimpleModelsCooling} for the evolution of the
\ac{ICM}, incorporating  additional heating
sources, and applying it to initial conditions based on observations of the
Milky Way~\citep{miller2016INTERACTIONFERMIBUBBLES} and stacked samples of Milky
Way-mass galaxies~\citep{zhang2024HotCircumgalacticMediuma}.
Our primary extensions to the \citet{kaiser2003SimpleModelsCooling} model are:
\begin{enumerate}
	\item a more detailed calculation of the cooling rate of the gas, which considers the incident photoionizing flux coming from the uniform \ac{UVB} and local sources in the Milky Way (namely \sgrAs, and star formation in the disc and in the \ac{CMZ}), and
	\item a mechanical heating term in the centre of the system by injecting thermal energy within a Gaussian kernel of a given length-scale $\lambda$ centred at $r=0$.
\end{enumerate}

We apply the model to two initial hot gas profiles: the first (\mbnormal) is based
on the observations from~\citet{miller2016INTERACTIONFERMIBUBBLES}, while the
second is a more massive, rescaled version of the former  (\mbrescaled) that is more in line with
recent eROSITA observations ~\citep{ponti2023AbundanceTemperatureOuter,zhang2024HotCircumgalacticMediuma}.
Our main results can be summarized as follows:
\begin{itemize}
	\item In the absence of any heating mechanism (photoionization or mechanical
	      feedback), the inner few \si{\kpc} of the hot \ac{CGM} cools rapidly
	      (in 22 to \SI{29}{Myr}). This cooling occurs on time-scales that
	      are much shorter than the time for which we expect the hot
	      \ac{CGM} to exist.

	\item Photoionization from the \ac{UVB} and local galactic sources do not
	      provide enough heating to stop the gas from cooling rapidly in the
	      centre, which implies additional heating mechanisms are needed to
	      offset radiative losses in the inner \ac{CGM}.

	\item In all model runs that do not include mechanical heating from a central
	      source, the gas at radii larger than a few \si{\kpc} cools on longer
	      time-scales compared to cooling in the centre, and flows inwards at subsonic
	      speeds. The mass inflow rates $\dot M(r)$ in this region are constant during
	      the evolution with values between $\dot M\sim 0.1-\SI{0.2}{}$ and
	      $0.6-\SI{0.7}{\Msolpyear}$ for the \mbnormal~and \mbrescaled~initial
	      conditions, respectively. Both values are too low to fully supply the Milky Way
	      disc with enough gas to sustain the current levels of star formation, but a
	      \ac{CGM} with mass similar to the \mbrescaled~\ac{CGM} could contribute significantly.

	\item At large radii, the mass inflow rates predicted by our model are in
	      close agreement with the self-similar solutions of the steady-state
	      cooling-flow model of \citet{stern2019CoolingFlowSolutions}, despite
	      the different assumptions underlying the two approaches.

	\item By varying the thermal energy that is injected into the centre of the
	      system, we find that the mechanical power needed to keep the \ac{CGM}
	      approximately in equilibrium, which we call the \textit{threshold
		      power}, is $10^{41-41.25}\,\si{\erg\per\s}$ for \mbnormal~and
	      $10^{41.5-41.75}\,\si{\erg\per\s}$ for \mbrescaled, due to its higher
	      initial density. The values for the threshold power decrease if the
	      spatial distribution of the injected energy is more concentrated.

	\item Assuming that the source that created the Fermi bubbles operates in duty
	      cycles, we estimate the average energy that is injected into the \ac{CGM}
	      for several different plausible astrophysical scenarios from the
	      literature. These average injected energies are comparable to or above
	      the threshold powers that we determined above. This indicates that the
	      source that created the Fermi bubbles is also powerful enough to stop the
	      cooling losses of the hot \ac{CGM} at the centre of the Milky Way.
	      Moreover, this similarity indicates that the same source could provide
	      enough energy to form a self-regulating cycle with the hot \ac{CGM} of
	      the Milky Way that would keep the \ac{CGM} stable for several \si{Gyr}.

	\item Whether sustained supernova feedback from the \ac{CMZ} can provide
	      enough energy to offset the rapid central cooling in the \ac{CGM} depends on
	      the initial condition and length-scale of the energy that is injected. This
	      mechanism fails to provide sufficient energy for the \mbrescaled~\ac{CGM},
	      but does provide enough energy for the \mbnormal~\ac{CGM} with energy
	      injection length-scale $\lambda=\SI{5}{\kpc}$. Supernovae from the disc
	      could increase the energy that is injected, though its exact contribution
	      is too uncertain to precisely quantify. Therefore, we cannot completely
	      rule out supernova feedback as a mechanism to stop the cooling in the
	      centre of the Milky Way's \ac{CGM}.

\end{itemize}

We have also shown that many of the above results remain unaffected when we
change the initial temperature distribution
(Section~\ref{sec:different-initial-condition}) or the length-scale of the kernel
that injects mechanical energy
(Sections~\ref{sec:length-scale},~\ref{sec:idealized-heating-case}).
Furthermore, we show that several assumptions underlying our evolutionary model (our treatment of convection and the choice of inner integration radius) do not
impact our conclusions, though we note that hydrostatic equilibrium itself
breaks down once the central region enters runaway cooling
(Section~\ref{sec:model-assumptions}).


\section*{Acknowledgements}

This work has received funding from the European Research Council (ERC) under
the Horizon Europe research and innovation programme (Acronym: FLOWS, Grant
number: 101096087). We thank an anonymous referee for their constructive
feedback. We thank Lucia Armillotta and the Cloudy-astrophysics forum
for help with Cloudy, and thank Gabriele Pezzulli, Joop Schaye, Marc Voit, Yi
Zhang and Bocheng Zhu for the productive discussions related to this work. This
work has made use of the Python programming language, in particular we
acknowledge the use of the
Astropy~\citep{theastropycollaboration2022AstropyProjectSustaining},
Numpy~\citep{harris2020ArrayProgrammingNumPy},
Scipy~\citep{virtanen2020SciPy10Fundamental},
Matplotlib~\citep{hunter2007Matplotlib2DGraphics} and
Numba~\citep{lam2015NumbaLLVMbasedPython} libraries. Large language models were
used for minor code autocompletion and for language feedback on the draft.

\section*{Data Availability}

All the code and scripts that have been used to produce the results in this work
can be found at a GitHub repository (\url{https://github.com/StenSipma/cgm-evolve}).
The various cooling tables that have been produced have also been
included in this repository.



\bibliographystyle{mnras}
\bibliography{bibtex} 



\appendix

\section{Supplementary figures}\label{app:figures}
\begin{figure*}
	\begin{center}
		\includegraphics{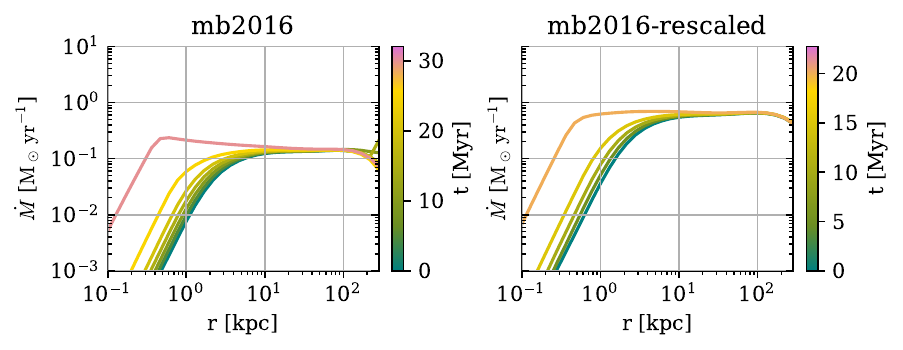}
	\end{center}
	\caption{
		The same as Figure~\ref{fig:results-compare-mass-accretion-cooling}, but
		with photoionization included.
	}\label{fig:results-compare-mass-accretion-pi}
\end{figure*}

\begin{figure*}
	\begin{center}
		\includegraphics{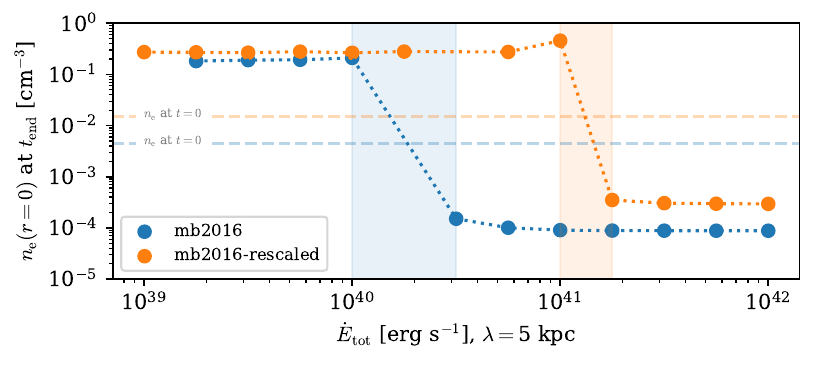}
	\end{center}
	\caption{
		The same as Figure~\ref{fig:final-density-S8}, but
		with $\lengthscale = \SI{5}{kpc} $.
	}\label{fig:final-density-S5}
\end{figure*}

\begin{figure*}
	\begin{center}
		\includegraphics{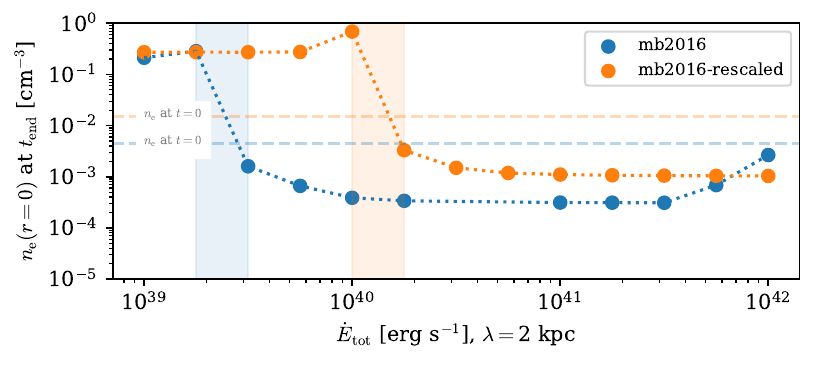}
	\end{center}
	\caption{
		The same as Figure~\ref{fig:final-density-S8}, but
		with $\lengthscale = \SI{2}{kpc}$.
	}\label{fig:final-density-S2}
\end{figure*}

\begin{figure*}
	\centering
	\begin{subfigure}{0.45\textwidth}
		\begin{center}
			\includegraphics{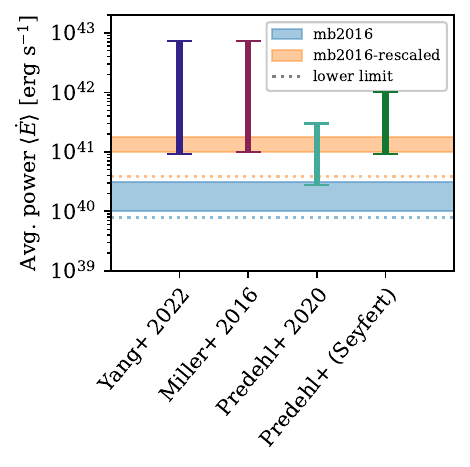}
		\end{center}
	\end{subfigure}%
	~
	\begin{subfigure}{0.45\textwidth}
		\begin{center}
			\includegraphics{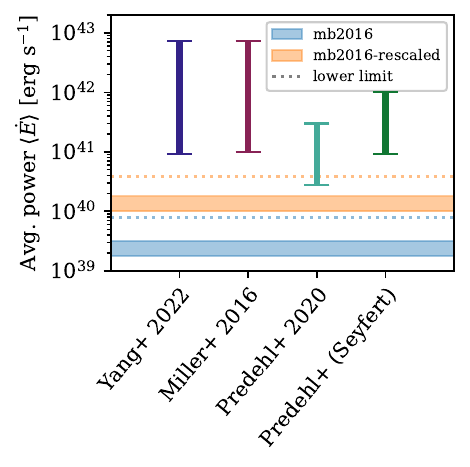}
		\end{center}
	\end{subfigure}

	\caption{
		The same as Figure~\ref{fig:threshold-comparison-S8}, but for $\lengthscale=\SI{5}{kpc}$ (left), and $\lengthscale=\SI{2}{kpc}$ (right).
	}\label{fig:threshold-comparison-S5-2}
\end{figure*}

This Appendix contains several figures that are briefly discussed in Section~\ref{sec:Results}. In particular, the effect of photoionization on the mass
inflow rate is shown in Figure~\ref{fig:results-compare-mass-accretion-pi}, while the remaining figures illustrate how varying the length-scale for mechanical energy injection affects the determination of threshold power values (Figs.~\ref{fig:final-density-S5} and \ref{fig:final-density-S2}) and
their comparison with several observational estimates of the average power required to generate the Fermi bubbles (Fig.~\ref{fig:threshold-comparison-S5-2}).

\begin{figure*}
	\begin{center}
		\includegraphics{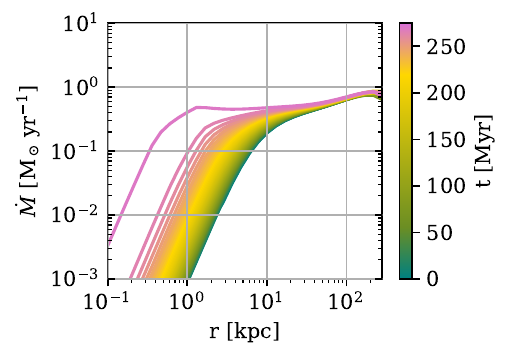}
	\end{center}
	\caption{
		The same as Figure~\ref{fig:results-compare-mass-accretion-cooling}, but
		for an initial condition with larger core radius $r_{\rm c}$ and lower temperature at the virial radius as described in Section~\ref{sec:different-initial-condition}.
	}\label{fig:discussion-mass-accretion-pi-zhang2024}
\end{figure*}


\bsp	
\label{lastpage}
\end{document}